%% file: main.tex
\PassOptionsToPackage{hyphens}{url}
\documentclass[sigconf,nonacm]{acmart}

\usepackage{graphicx}
\usepackage{bm}
\usepackage{wrapfig}

\input{macros}

\begin{document}

\title{A Shape-Adaptive Architecture with Disaggregated Quantization for Efficient LLM Serving}

\author{Cong Guo}
\authornote{Both authors contributed equally to this research.}
\affiliation{%
  \institution{Duke University}
  \city{Durham}
  \state{NC}
  \country{USA}}
\email{cong.guo@duke.edu}

\author{Chiyue Wei}
\authornotemark[1]
\affiliation{%
  \institution{Duke University}
  \city{Durham}
  \state{NC}
  \country{USA}}
\email{chiyue.wei@duke.edu}

\author{Bowen Duan}
\affiliation{%
  \institution{Duke University}
  \city{Durham}
  \state{NC}
  \country{USA}}
\email{bowen.duan@duke.edu}

\author{Haoxuan Shan}
\affiliation{%
  \institution{Duke University}
  \city{Durham}
  \state{NC}
  \country{USA}}
\email{haoxuan.shan@duke.edu}

\author{Benjamin F. Morris III}
\affiliation{%
  \institution{Duke University}
  \city{Durham}
  \state{NC}
  \country{USA}}
\email{ben.morris@duke.edu}

\author{Yintao He}
\affiliation{%
  \institution{Duke University}
  \city{Durham}
  \state{NC}
  \country{USA}}
\email{yintao.he@duke.edu}

\author{Hai ``Helen'' Li}
\affiliation{%
  \institution{Duke University}
  \city{Durham}
  \state{NC}
  \country{USA}}
\email{hai.li@duke.edu}

\author{Yiran Chen}
\affiliation{%
  \institution{Duke University}
  \city{Durham}
  \state{NC}
  \country{USA}}
\email{yiran.chen@duke.edu}

\input{content/0.abstract}

\maketitle

\input{content/1.intro}
\input{content/2.background}
\input{content/3.dynacore}
\input{content/4.decoding}
\input{content/5.prefill}
\input{content/5.5.scheduling}
\input{content/6.evaluation}
\input{content/7.conclusion}

\bibliographystyle{ACM-Reference-Format}
\bibliography{refs}

\end{document}

%% file: macros.tex
\usepackage{xspace}
\usepackage{multirow}
\usepackage{booktabs}
\usepackage[ruled,vlined,linesnumbered,algo2e]{algorithm2e}

\newcommand{\dyna}{\textit{DynaCore}\xspace}

\newcommand{\Fig}[1]{Fig.~\ref{#1}}
\newcommand{\Tbl}[1]{Tbl.~\ref{#1}}
\newcommand{\Alg}[1]{Alg.~\ref{#1}}

%% file: content/0.abstract.tex
\begin{abstract}
Large language models (LLMs) have become the backbone of modern AI applications, but pose significant challenges for efficient inference.
Their autoregressive generation divides execution into two phases: \textit{prefill}, dominated by large GEMMs, and \textit{decoding}, dominated by small GEMVs.
Modern serving systems further introduce complexity through continuous batching and prefill--decoding disaggregation, leading to dynamic workloads and phase separation.
However, existing accelerators remain poorly aligned with these system-level behaviors, resulting in inefficiencies in LLM serving.

In this work, we present \textbf{\dyna{}}, a unified architecture for efficient LLM serving via system--architecture co-design.
We observe that the compute tile a systolic array executes, its \textit{Minimum Efficient Unit} (MEU), spans all three GEMM dimensions.
\dyna{} reshapes the MEU along all three: spatially it trades array width against height \textit{asymmetrically}, raising weight delivery while leaving the input path untouched, and temporally \textit{Split-$K$} maps the reduction onto the array, folding partial sums through the interconnect the array already has.
To exploit phase separation, we further propose \textit{disaggregated quantization}, applying dual-side quantization to prefill and weight-only quantization to decoding, with an inner-product mixed-precision datapath that keeps output width invariant to precision.
A \textit{runtime scheduling framework} then selects an MEU per batch.
Evaluation with real-world serving traces shows that \dyna{} substantially reduces service-level latency over quantization and reconfigurable accelerators, improving \textit{TTFT} by $3.50\times$ and $2.97\times$ and \textit{TPOT} by $36.55\times$ and $8.02\times$, respectively.
\end{abstract}

%% file: content/1.intro.tex
\section{Introduction}
Large Language Models (LLMs)~\cite{Devlin2018BERT,hugo2023llama,touvron2023llama2,llama2024llama3,radford2019language_gpt2,dettmers2022gpt3,jiang2023mistral_7b,chen2025dpad} have become the foundation of a wide range of real-world applications, including natural language understanding~\cite{touvron2023llama2,deepseek2025deepseek_r1,OpenAI2023GPT4,zhu2024multilingual} and code generation~\cite{thakur2024verigen,xiong2025hlspilot,zhuo2024bigcodebench}.
As model sizes continue to grow, LLM serving increasingly stresses computing infrastructure, with inference emerging as the primary performance and cost bottleneck.
A key challenge in LLM inference stems from its autoregressive generation paradigm~\cite{vaswani2017attention,katharopoulos2020transformers_are_rnns,wang2020linformer}, which divides execution into two fundamentally different phases: \textit{prefill} and \textit{decoding}~\cite{pope2023efficiently}.
The prefill phase processes all input tokens in parallel using compute-intensive matrix--matrix multiplications (GEMMs), achieving high utilization on modern accelerators.
In contrast, the decoding phase generates tokens sequentially, where each step is dominated by matrix--vector multiplications (GEMVs) with limited parallelism and low arithmetic intensity.
This fundamental shift from throughput-oriented to latency-sensitive computation leads to severe hardware underutilization during decoding, making it the dominant bottleneck in end-to-end LLM serving~\cite{kwon2023efficient_vllm,agrawal2023sarathi,patel2024splitwise,pope2023efficiently}.

To mitigate the inefficiency of the decoding phase, prior works have explored optimizations across both algorithmic and system levels.
From an algorithmic perspective, decoding is widely recognized as a memory-bound workload in which weight accesses dominate execution (often accounting for over 90\% of memory traffic), while activation footprints remain relatively small~\cite{pope2023efficiently, holmes2024deepspeed}.
This observation has motivated weight-only quantization techniques~\cite{lin2024awq,kim2023squeezellm,shao2023omniquant}, which reduce memory bandwidth demand by compressing model weights, thereby improving performance in memory-bound regimes.
From a system perspective, continuous batching (CB)~\cite{yu2022orca,holmes2024deepspeed} improves effective parallelism by aggregating decoding requests.
Additionally, when prefill and decoding are co-located on the same device, resource contention leads to significant latency variability.
Prefill–decoding disaggregation (PDD)~\cite{patel2024splitwise,zhong2024distserve,qin2024mooncake} alleviates this interference by separating the two phases across devices, reducing decoding latency.

Despite these advances, existing accelerator designs remain poorly aligned with the combined effects of algorithmic and system-level optimizations in LLM serving.
This misalignment arises because prefill and decoding exhibit fundamentally different execution characteristics, yet current designs lack cross-layer co-design across algorithms, systems, and hardware.

First, decoding remains a fundamental bottleneck due to the low hardware utilization of GEMV operations.
Modern accelerators, particularly systolic-array-based designs~\cite{kung1982systolic,chen2016eyeriss,samajdar2018scale,jouppi2017datacenter,jouppi2023tpu}, are optimized for large GEMMs and exhibit poor efficiency when processing narrow GEMVs.
While continuous batching (CB) partially increases parallelism, it introduces highly dynamic GEMM shapes that depend on the number of active tokens, leading to fluctuating execution patterns.
As a result, CB does not resolve the architectural mismatch between decoding workloads and hardware execution models (Section~\ref{subsec:array}).
Achieving high efficiency, therefore, requires architectures that better support fine-grained parallelism and align with decoding-specific computation patterns.

Second, quantization~\cite{lin2024awq,lin2024qserve,frantar2022gptq,shao2023omniquant,dettmers2022gpt3,kim2023squeezellm,xiao2023smoothquant,ashkboos2024quarot}, a key technique for improving LLM efficiency, is usually applied with an identical configuration (e.g., W8A8 or W4A16, where W$m$A$n$ denotes $m$-bit weights and $n$-bit activations) across both prefill and decoding phases.
However, these two stages exhibit fundamentally different bottlenecks, making a unified design inherently suboptimal (Section~\ref{subsec:quant}).
Dual-side quantization~\cite{xiao2023smoothquant,lin2024qserve} (W8A8) improves compute efficiency and maintains high accuracy during prefill but offers limited benefit for decoding, where performance is constrained by memory bandwidth.
In contrast, weight-only quantization~\cite{lin2024awq,kim2023squeezellm} (W4A16) reduces memory traffic during decoding but provides little acceleration for prefill and degrades accuracy.
This mismatch highlights the need for phase-specific quantization that adapts to the distinct compute and memory characteristics of each stage.

Finally, these challenges highlight the need for a unified system--architecture co-design.
Deploying separate hardware for prefill and decoding leads to resource imbalance under dynamic PDD scheduling, while static architectures cannot adapt to CB-induced variability.
An effective solution must therefore support both phases within a unified architecture that can dynamically reconfigure its execution behavior and coordinate with system-level scheduling.

In this work, we present \textbf{\dyna{}}, a phase-adaptive architecture for efficient LLM serving.
Our starting point is that the compute tile a systolic array executes, which we call the \emph{Minimum Efficient Unit} (MEU), spans all three dimensions of a GEMM: $M$ and $N$ are partitioned \emph{spatially} across the array as the extents $m$ and $n$, while the reduction dimension $K$ is partitioned \emph{temporally} as the extent $k$.

First, we design a \textbf{\textit{splittable systolic array}} that reshapes the array spatially, trading $m$ against $n$ at a fixed PE count.
The reshaping is \emph{asymmetric}: flattening demands more weight bandwidth while leaving the input path unchanged, matching a workload that has little activation data to move but must stream weights to every PE.
Second, we introduce \textbf{\textit{Split-$K$}}, which partitions the reduction across the array once both spatial extents are pinned by the workload.
Since slices of $K$ must be summed rather than drained independently, we fold them with a \emph{tick--tock} schedule that alternates forwarding and accumulating along the links the array already has, adding no reduction network.

Third, we propose \textbf{\textit{disaggregated quantization}}, which supplies what the first two dimensions demand.
Prefill adopts dual-side quantization (W8A8) while decoding adopts weight-only quantization (W4A16), relieving the weight path that asymmetric reshaping stresses; its mixed-precision datapath directs low-precision throughput along $K$ rather than into additional outputs, leaving every array configuration reachable at either precision.
Finally, we develop a \textbf{\textit{runtime scheduling framework}} that selects an MEU for each batch, searching the reachable configurations against a joint latency--energy objective from offline profiling.

\dyna{} integrates these techniques into a unified design for practical LLM serving under dynamic workloads.
Comprehensive evaluation demonstrates significant gains in throughput, energy efficiency, and latency metrics (\textit{time-to-first-token (TTFT)} and \textit{time-per-output-token (TPOT)}) over state-of-the-art LLM accelerators.
In summary, our main contributions are as follows:
\begin{itemize}
    \item We design a \emph{splittable systolic array} that reshapes the MEU along all three dimensions, two spatially and one temporally, without altering the array topology.

    \item We propose \emph{disaggregated quantization} with an inner-product mixed-precision datapath, supplying the weight bandwidth that reshaping demands.

    \item We develop a \emph{runtime scheduling framework} that selects an MEU per batch, bridging system-level dynamics and architectural adaptation under continuous batching and disaggregated serving.

    \item Comprehensive evaluation shows that \dyna{} substantially reduces service-level latency over the FIGLUT~\cite{park2025figlut} quantization accelerator and the Planaria~\cite{ghodrati2020planaria} reconfigurable accelerator, improving \textit{TTFT} by \textbf{3.50$\times$} and \textbf{2.97$\times$} and \textit{TPOT} by \textbf{36.55$\times$} and \textbf{8.02$\times$}, respectively, under real serving traces.
\end{itemize}

%% file: content/2.background.tex
\section{Background and Motivation}
\label{sec:motivation}

\begin{figure}[t]
    \centering
    \includegraphics[width=1\linewidth]{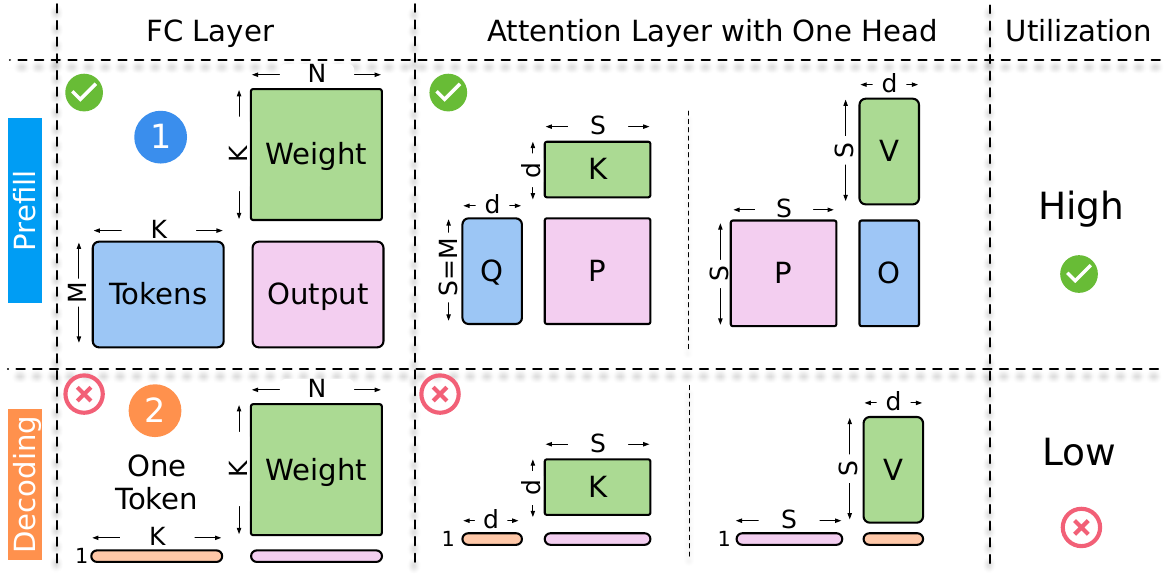}
    \vspace{-5mm}
    \caption{
    GEMM characteristics across prefill and decoding phases in LLM.}
    \vspace{-6mm}
    \label{fig:gemm}
\end{figure}

\subsection{LLM Inference and Characteristics}
Large Language Models (LLMs)~\cite{Devlin2018BERT,jiang2023mistral_7b,chen2025dpad}, such as GPT~\cite{radford2018improving,radford2019language_gpt2,dettmers2022gpt3,achiam2023gpt} and LLaMA~\cite{hugo2023llama,touvron2023llama2,grattafiori2024llama}, are built on the Transformer architecture~\cite{vaswani2017attention}, comprising stacked attention and feed-forward layers.
During inference, LLMs generate text in an \textit{autoregressive} manner.

\textbf{Prefill.}
Given a prompt, the model first performs a {prefill} phase, processing all input tokens in parallel across Transformer blocks.
As shown in Figure~\ref{fig:gemm}, this phase primarily involves large matrix-matrix multiplications (GEMMs), which are well-suited to the dense compute structures of modern accelerators.

\textbf{Decoding.}
After the prefill phase, the model enters the {decoding} phase, where tokens are generated sequentially, one at a time.
As shown in Figure~\ref{fig:gemm}, each iteration reuses the same model weights while processing only a single token, turning the computation into a matrix-vector multiplication (GEMV) with extremely limited parallelism and data reuse.
Although both phases share the same model architecture, prefill is compute-bound, whereas decoding is memory-bound and severely underutilizes compute resources~\cite{kwon2023efficient_vllm,agrawal2023sarathi,patel2024splitwise}.

\textbf{Takeaway:}
Prefill produces large, highly parallel GEMMs, whereas decoding produces GEMMs whose token dimension collapses to $M{=}1$.
This intrinsic mismatch leads to substantial efficiency gaps on conventional accelerators.

\subsection{Rethinking System Scheduling}
\label{subsec:system}

Modern LLM serving systems~\cite{kwon2023efficient_vllm,yu2022orca,zheng2024sglang} introduce highly dynamic workloads driven by runtime scheduling, which continuously changes the balance between prefill and decoding.
These dynamics expose a fundamental mismatch between modern LLM serving and static hardware execution models.

\textbf{Continuous Batching (CB).}
CB~\cite{yu2022orca,kwon2023efficient_vllm,agrawal2023sarathi} aggregates requests to improve decoding throughput, but yields highly variable GEMM shapes depending on the number of active tokens.
Batch sizes fluctuate as requests arrive and complete, so computation shifts continuously between flat, memory-bound GEMVs and larger, more parallel workloads, leaving hardware utilization inconsistent.

\textbf{Prefill-Decoding Disaggregation (PDD).}
To mitigate interference between phases, modern systems adopt PDD~\cite{patel2024splitwise,qin2024mooncake,zhong2024distserve}, which executes the two phases on separate devices and thereby removes the resource contention that degrades decoding throughput when they are co-located.
While PDD improves performance, it also exposes a key limitation: statically deploying separate hardware for prefill and decoding leads to resource imbalance under dynamic workloads, as the ratio between phases continuously changes.

\textbf{Takeaway.}
LLM serving exhibits highly dynamic workload characteristics, where both GEMM shapes and phase ratios vary over time.
This requires a unified architecture that can efficiently support both phases and adapt to changing workloads.
Such adaptivity must be coordinated with system scheduling, enabling joint system–architecture co-design across quantization, array organization, and execution scheduling.

\begin{figure}[t]
    \centering
    \includegraphics[width=0.95\linewidth]{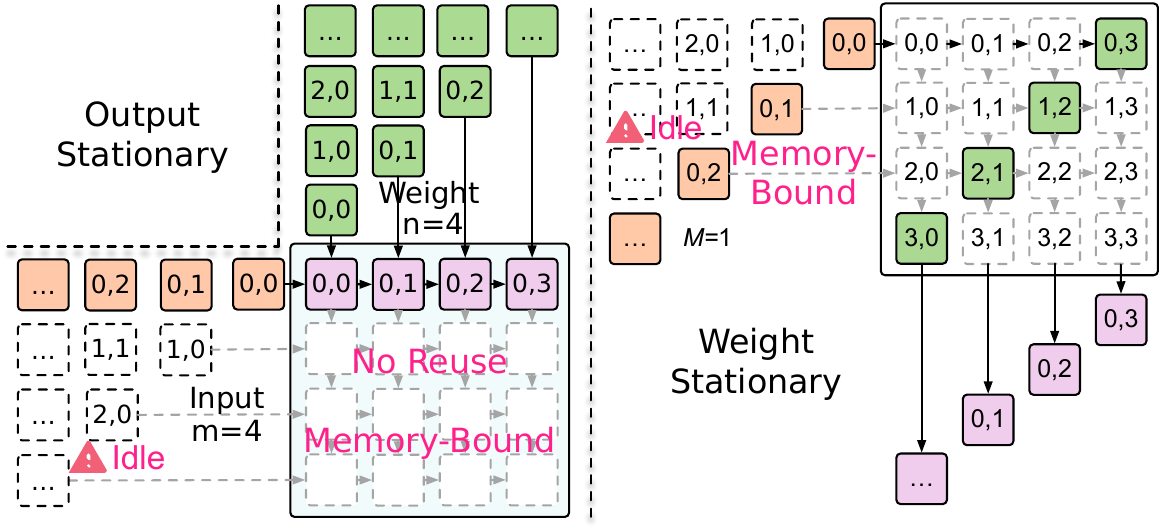}
    \vspace{-4mm}
\caption{
Systolic array inefficiency under decoding ($M{=}1$) for both output-stationary (left) and weight-stationary (right) dataflows.
}
    \label{fig:array}
    \vspace{-5mm}
\end{figure}

\subsection{Rethinking Array Architecture}
\label{subsec:array}
Systolic-array–based accelerators~\cite{kung1982systolic,chen2016eyeriss,samajdar2018scale,jouppi2017datacenter,jouppi2023tpu} are widely used for LLM inference due to their high throughput and efficient data reuse.
However, their efficiency relies on GEMM patterns, which break down under decoding GEMV and are further exacerbated by dynamic GEMM shapes introduced by CB.

\textbf{Notation.}
Throughout this paper, uppercase $M \times N \times K$ denotes the full dimensions of a GEMM as issued by the model, while lowercase $m \times n \times k$ denotes the tile the hardware executes as one unit.

\textbf{Minimum Efficient Unit.}
To analyze the efficiency limits of such architectures under LLM workloads, we introduce the \textbf{\textit{Minimum Efficient Unit} (MEU)}: the smallest GEMM tile that fully utilizes both the compute array and its on-chip data movement.
An MEU is inherently \textbf{\emph{three-dimensional}}, $\mathrm{MEU} = m \times n \times k$.
The spatial extents $m$ and $n$ are set by the physical array organization and bound the output tile produced in parallel.
The third extent $k$ is temporal: it is the reduction depth the array must consume before every PE becomes active.
Unlike $m$ and $n$, $k$ is not fixed by the array dimensions alone; it depends on how operands are delivered and propagated, and must be determined per implementation.
Utilization for a GEMM of shape $M \times N \times K$ is therefore
\begin{equation}
U \;=\; \underbrace{\frac{\min(M,m)\cdot\min(N,n)}{m\cdot n}}_{\text{spatial}} \;\cdot\; \underbrace{\frac{K}{K+k}}_{\text{temporal}} .
\label{eq:meu}
\end{equation}
Full utilization requires $M \ge m$, $N \ge n$, \emph{and} $K \gg k$ simultaneously.
We call a GEMM meeting all three conditions a \textbf{tall GEMM}, and one with $M < m$ a \textbf{flat GEMM}.
Decoding produces flat GEMMs: with $M{=}1$ on an $S \times S$ array the spatial term collapses to $1/S$.

\begin{table}[t]
\centering
\caption{Representative accelerators and their MEU adaptability.
\dyna{} figures are per core.}
\vspace{-3mm}
\label{tab:llm_accel_summary}
\resizebox{\linewidth}{!}{
\begin{tabular}{lcccc}
\toprule
\textbf{Architecture} & \textbf{MEU ($m\times n$)} & \textbf{Dims} & \textbf{Adaptation} & \textbf{Target} \\
\midrule
\multicolumn{5}{l}{\textit{Reconfigurable arrays: conformal fission}} \\
Planaria~\cite{ghodrati2020planaria} & 128$\times$128 $\to$ 32$\times$32 & $m{=}n$ & Conformal & Multi-tenant DNN \\
SARA~\cite{samajdar2022self} & 128$\times$128 $\to$ 4$\times$4 & $m{=}n$ & Conformal & Multi-DNN scaling \\
\midrule
\multicolumn{5}{l}{\textit{Fixed arrays: LLM quantization / sparsity accelerators}} \\
Olive~\cite{guo2023olive} & 32$\times$32 & --- & Fixed & Prefill GEMM \\
BitVert~\cite{chen2024bbs} & 16$\times$32 & --- & Fixed & Prefill GEMM \\
FIGNA~\cite{jang2024figna} & 128$\times$128 & --- & Fixed & Prefill GEMM \\
FIGLUT~\cite{park2025figlut} & 64$\times$64 & --- & Fixed & Prefill GEMM \\
TransArray~\cite{guo2025transitive} & 192$\times$256 & --- & Fixed & Prefill GEMM \\
\midrule
\multicolumn{5}{l}{\textit{Reconfigurable arrays: asymmetric reshaping}} \\
\textbf{\dyna{}} (Ours) & 32$\times$32 $\to$ 1$\times$1024 & $\bm{m,n,k}$ & \textbf{Asymmetric} & \textbf{Prefill + Decoding} \\
\bottomrule
\end{tabular}
}
\vspace{-3mm}
\end{table}

\textbf{Systolic Array Limitation.}
Modern LLM accelerators~\cite{guo2023olive,jang2024figna,guo2025transitive,park2025figlut} predominantly adopt 2D systolic arrays to exploit data reuse across both spatial dimensions, amortizing on-chip buffer bandwidth.
Table~\ref{tab:llm_accel_summary} summarizes representative designs and the MEU each of them realizes.
However, as illustrated in Figure~\ref{fig:array}, decoding fundamentally breaks this reuse pattern.
With only a single input token ($M{=}1$), activations cannot be reused across rows, and weights cannot be efficiently shared across the array.
As a result, reuse collapses, leaving most processing elements idle.
This reveals a fundamental mismatch: systolic arrays are designed to trade on-chip bandwidth for data reuse, while decoding workloads provide little opportunity for reuse.
Consequently, existing designs fail to efficiently support decoding, despite being well optimized for prefill.

\textbf{Prior Reconfigurable Arrays.}
Reconfigurable arrays are not new, but existing designs reshape only the spatial pair $(m,n)$, and do so \emph{symmetrically}.
Planaria~\cite{ghodrati2020planaria} fissions a large array into fixed square sub-arrays so that concurrent DNNs can be co-located; SARA~\cite{samajdar2022self} learns flexible GEMM configurations and scales a monolithic array down to finer square cells to match varying layer dimensions.
Both target \emph{multi-tenant or multi-DNN} execution, where each tenant still presents a well-formed 2D tile, and both therefore traverse only the conformal diagonal $m = n$, as listed in Table~\ref{tab:llm_accel_summary}.
This constraint, not the granularity of the fission, is what excludes decoding.
Conformal splitting shrinks $m$ and $n$ together, whereas a flat GEMM requires them to move in \emph{opposite} directions: $m$ must fall toward one while $n$ grows to absorb the freed PEs.
No amount of finer conformal subdivision reaches this operating point, which is why Planaria and SARA saturate at MEU floors of $32{\times}32$ and $4{\times}4$ respectively.

\textbf{LLM-Related Accelerators.}
Recent LLM accelerators~\cite{guo2022ant,guo2023olive,jang2024figna,chen2025bitmod,chen2024bbs,hu2025mant,guo2025transitive,park2025figlut} improve efficiency through quantization, sparsity, or lookup-based computation, but keep a fixed array organization throughout execution.
Their MEUs, listed in Table~\ref{tab:llm_accel_summary}, are static and in several cases far larger than a decoding GEMV.
While these techniques reduce arithmetic and memory cost, they all assume efficient data reuse and leave the array shape untouched, so none addresses decoding underutilization.

\textbf{Takeaway.}
Decoding is not merely a \emph{small} workload; it is \emph{dimensionally mismatched}.
This motivates a \textit{splittable systolic array} that reshapes its MEU across space and time and adapts its operand paths asymmetrically, listed in Table~\ref{tab:llm_accel_summary}.

\begin{figure}[t]
    \centering
    \includegraphics[width=0.95\linewidth]{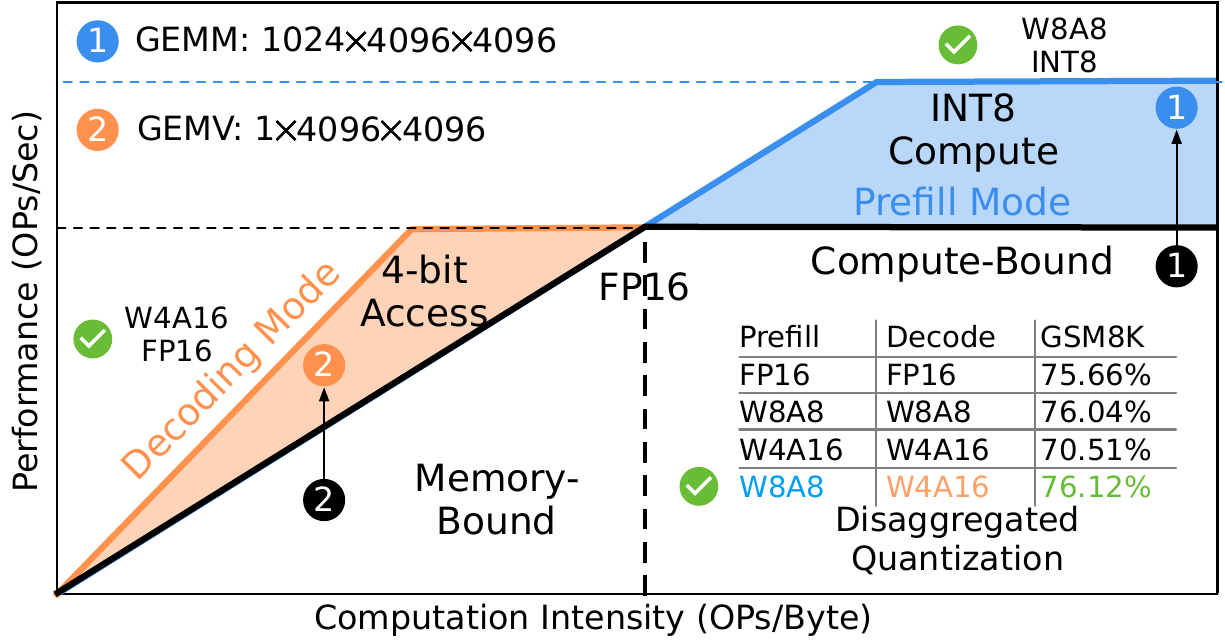}
\vspace{-4mm}
\caption{
Overview of disaggregated quantization.
Roofline analysis illustrates the contrast between GEMM (prefill) and GEMV (decoding).
Bottom-right: accuracy comparison on LLaMA3-8B evaluated on GSM8K.
}
\label{fig:dq}
\vspace{-8pt}
\end{figure}

\subsection{Rethinking Quantization Strategies}
\label{subsec:quant}
Quantization~\cite{lin2024awq,lin2024qserve,frantar2022gptq,shao2023omniquant,dettmers2022gpt3,kim2023squeezellm,xiao2023smoothquant,ashkboos2024quarot} improves LLM inference efficiency by reducing the precision of weights and activations.
Existing approaches typically adopt a unified configuration across both phases, as reflected in prior accelerators in Table~\ref{tab:llm_accel_summary}.
However, as the roofline in Figure~\ref{fig:dq} shows, prefill and decoding sit on opposite sides of the ridge point, making a single scheme suboptimal.

\textbf{Decoding.}
Operating in the memory-bound region, decoding is dominated by weight traffic, so weight-only quantization~\cite{lin2024awq,kim2023squeezellm} (e.g., W4A16) reduces the memory-bound latency floor in proportion to the weight bit-width.
Activations, whose traffic is negligible, are best left in FP16: this preserves accuracy at little memory cost, whereas dual-side quantization~\cite{lin2024qserve,xiao2023smoothquant,ashkboos2024quarot} (e.g., W8A8) raises weight precision and thus traffic.
Reducing compute precision brings little benefit either, since decoding demands far less arithmetic than FP16 or INT8 units can supply.
This gain, however, materializes only if the compute array keeps pace with the memory system.
A GEMV on an $S{\times}S$ output-stationary array activates a single row, leaving $1/S$ of the PEs busy; on a $128{\times}128$ array at 500\,MHz with 4\,TB/s of bandwidth, the resulting compute time exceeds the FP16 weight-streaming time by about $31\times$, regardless of model size.
Decoding on such an array is therefore utilization-bound rather than bandwidth-bound, and weight-only quantization pays off only after array utilization is restored (Section~\ref{sec:array}).

\textbf{Prefill.}
For long prompts, prefill sits far from the memory-bound region, so reducing memory traffic yields little.
Here throughput is the constraint, and dual-side quantization (e.g., W8A8) is the better fit, doubling arithmetic throughput through INT8 units while maintaining model accuracy.
Prefill GEMMs are tall only when the prompt is long, however: since $M$ equals the number of prompt tokens, a short prompt yields $M$ below the array dimension and inherits the same row underutilization as decoding.

\textbf{No Duplication Cost.}
Assigning different precisions to the two phases does not imply storing two copies of the model.
Under PDD, prefill and decoding already execute on separate devices, each holding only the weights for its own phase, so the W8A8 and W4A16 copies reside where they are used and neither device carries the other's format.

\textbf{Accuracy Gain.}
As shown in Figure~\ref{fig:dq} (bottom-right), weight-only quantization (e.g., W4A16) leads to noticeable accuracy degradation.
In contrast, disaggregated quantization, using W4A16 for decoding and W8A8 for prefill, achieves near-lossless accuracy while maintaining high efficiency.
Notably, this configuration outperforms both standalone W8A8 and W4A16, achieving a superior efficiency–accuracy trade-off.

\textbf{Takeaway.}
\textit{Disaggregated quantization} lowers precision on exactly the path that asymmetric reshaping stresses, leaving activations untouched on the path that needs no additional bandwidth.
Supporting both precisions in one datapath, however, requires a mixed-precision design aware of the array shape, one that aligns with the temporal dimension $k$ as well as with the asymmetric provisioning of operand paths.

%% file: content/3.dynacore.tex
\section{\dyna{} Architecture}
\label{sec:arch}

\begin{figure}[t]
    \centering
    \includegraphics[width=0.95\linewidth]{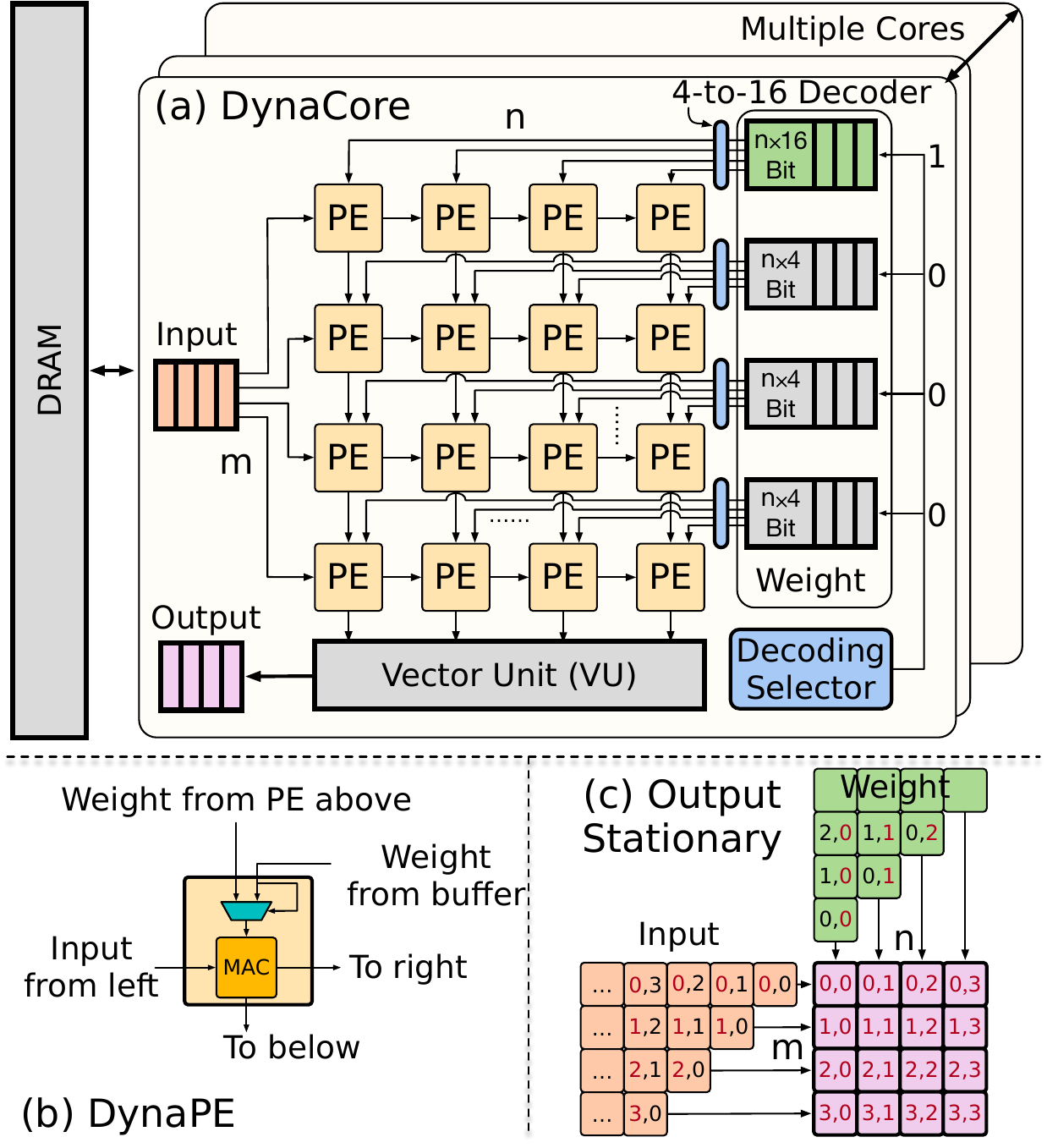}
    \vspace{-3mm}
    \caption{
    Overview of the \dyna{} architecture.
    (a) Multiple identical cores, each a splittable systolic array.
    (b) \textit{DynaPE} selects its weight source between the PE above and the buffer.
    (c) Output-stationary dataflow.
    }
    \vspace{-6mm}
    \label{fig:main_arch}
\end{figure}

\dyna{} is organized as a set of identical cores, each an output-stationary systolic array with its own input, weight, and output buffers, as shown in Figure~\ref{fig:main_arch}(a).
A core is the unit at which an MEU is realized, and cores reshape independently, so a device can run several MEU configurations concurrently.

\begin{figure*}[t]
    \centering
    \includegraphics[width=0.95\linewidth]{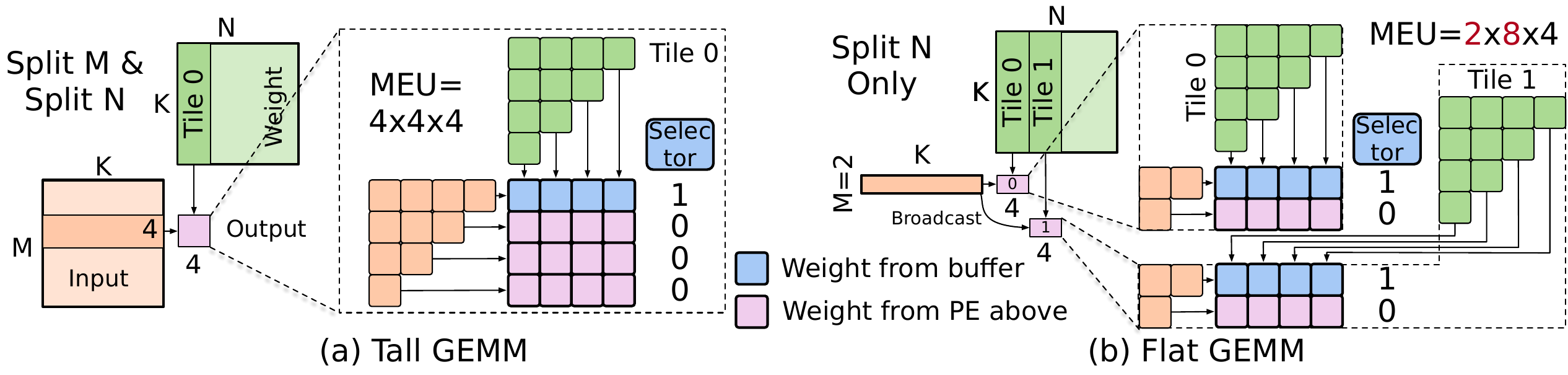}
    \vspace{-2mm}
\caption{
Spatial reshaping in \dyna{}.
(a) A tall GEMM is split along both $M$ and $N$, and the core runs as a square array.
(b) A flat GEMM is split along $N$ only; the core halves $m$ and doubles $n$, broadcasting the input while doubling weight delivery.
}
    \label{fig:mode}
    \vspace{-4mm}
\end{figure*}

Within a core, reconfiguration proceeds along two axes.
\emph{Spatially}, the array trades $m$ against $n$ at a fixed PE count.
The mechanism is the selector inside each \textit{DynaPE} (Figure~\ref{fig:main_arch}(b)), which chooses whether the PE receives its weight from the PE above or directly from the buffer.
Rows fed from the buffer begin a new logical row of the reshaped array, so the selector pattern alone determines the $m \times n$ organization, leaving the array topology untouched.
Because only the weight side changes while the input side continues to broadcast a single stream, the two operand paths are provisioned asymmetrically rather than scaled together.
\emph{Temporally}, the reduction dimension becomes mappable: the array partitions a long reduction across PE groups and folds their partial sums in place, which the output-stationary dataflow (Figure~\ref{fig:main_arch}(c)) supports without widening the output path.

The remainder of this section follows the two axes in turn.
Section~\ref{sec:array} covers spatial reshaping, from tall GEMMs where a square configuration suffices to flat GEMMs where the weight and input paths must be provisioned asymmetrically.
Section~\ref{sec:splitk} adds the temporal axis through Split-$K$.
Section~\ref{sec:quantization} then shows how disaggregated quantization matches all three dimensions: it supplies the weight bandwidth that asymmetric reshaping of $m$ and $n$ demands, and its mixed-precision datapath directs the throughput freed at low precision along $K$.

%% file: content/4.decoding.tex
\section{Asymmetric Spatial Reshaping}
\label{sec:array}

This section describes how a core reshapes the spatial extents of its MEU.
The PE count is fixed at $S^2$, so $m$ and $n$ can only be traded against each other; what varies is how the trade is made and what it costs on each operand path.

\subsection{Selector-Based Reshaping}

Each \textit{DynaPE} chooses its weight source between the PE above and the weight buffer (Figure~\ref{fig:main_arch}(b)).
A PE fed from the buffer starts a new weight stream; one fed from above continues and reuses the stream of the row above it.
The selector pattern therefore partitions the array into groups, each forming one logical row of the reshaped array, so a pattern with $g$ asserted rows yields
\begin{equation}
m = S/g, \qquad n = g\,S, \qquad m \cdot n = S^2 .
\label{eq:reshape}
\end{equation}
Reshaping rewrites $S$ control bits rather than the interconnect, leaving the array topology unchanged, and its cost falls on one side only: $g$ groups require $g$ concurrent weight streams while the input side carries a single broadcast stream.

\subsection{Tall GEMM: Splitting $M$ and $N$}
\label{sec:tall}
A tall GEMM exceeds the core in both spatial dimensions, so it is tiled along $M$ and $N$ and each tile maps directly onto the array.
Figure~\ref{fig:mode}(a) shows the resulting configuration on a $4\times4$ core with a single asserted row.
Weights enter at the top and propagate downward, reused across all four rows; inputs propagate horizontally, reused across all four columns.
With $M \ge m$ and $N \ge n$, the spatial term of Equation~\ref{eq:meu} reaches unity, and reuse in both directions holds on-chip traffic at its minimum of $m{+}n$ operands per cycle for $m \cdot n$ PEs.
The temporal extent follows from the same mapping: an operand entering the array reaches the farthest PE after $\max(m,n)$ cycles, so the array saturates in four cycles, giving $\mathbf{MEU = 4\times4\times4}$.

\subsection{Flat GEMM: Asymmetric Regrouping}
\label{sec:flat}

\textbf{Reshaping the MEU.}
A flat GEMM leaves $M$ below the array dimension, so the tiling reduces to splitting $N$ alone: there is nothing left to split along $M$, and a square mapping simply wastes the rows it cannot fill.
The core instead reduces $m$ to the available token count and regroups the freed rows into wider logical rows along $N$.
Figure~\ref{fig:mode}(b) shows $M{=}2$ on the same core: two asserted rows, giving $m{=}2$ and $n{=}8$.
Although $m$ has halved, the temporal extent is unaffected.
Each group still spans four columns of the physical array, so an operand still reaches the farthest PE after $\max(m,n)=4$ cycles within its group, leaving $k{=}4$ and $\mathbf{MEU = 2\times8\times4}$.

\textbf{Asymmetric Adaptation.}
The two operand paths move in opposite directions under this regrouping.
Halving $m$ removes half of the vertical weight reuse, so the buffer must issue two independent weight streams instead of one; in general, $g$ groups require $g$ streams.
The input path is untouched, since the same tokens serve every tile and are simply broadcast.
This asymmetry is intrinsic to LLM decoding rather than an artifact of the mapping: with $M$ small there is little activation data to begin with, so widening the input path buys nothing, and all of the pressure falls on weight delivery.
It is the same observation that makes weight-only quantization the prevailing choice for decoding, and Section~\ref{sec:quantization} shows how the two meet.

\textbf{Continuum of Configurations.}
The square and flattened configurations are endpoints of a continuum rather than separate designs.
Any $g$ dividing $S$ is a valid selector pattern, so by Equation~\ref{eq:reshape} an $S \times S$ core realizes $S\times S$, $\tfrac{S}{2}\times 2S$, $\ldots$, $1 \times S^2$ at the same PE count; on a $32\times32$ core this gives six reachable shapes between fully square and fully flattened.
The temporal extent is invariant across all of them, since propagation is bounded by the physical array dimension regardless of how rows are grouped.
The array therefore tracks the token count of the current batch, which matters under continuous batching where that count changes from iteration to iteration.

\begin{figure}[t]
    \centering
    \includegraphics[width=\linewidth]{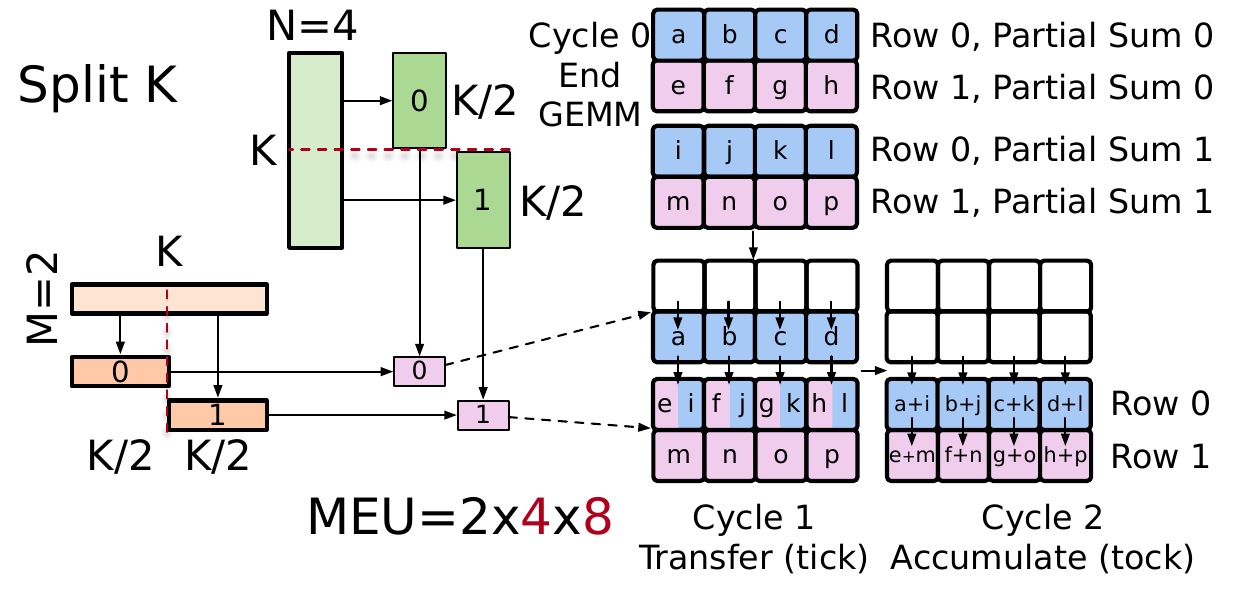}
    \vspace{-6mm}
\caption{
Temporal split-$k$ mapping.
The array is partitioned into $t=2$ groups along $K$, each producing a partial sum for the same output tile.
}
    \label{fig:splitk}
    \vspace{-4mm}
\end{figure}

\section{Temporal Split-$K$ Mapping}
\label{sec:splitk}
Spatial regrouping converts rows freed from $m$ into a wider $n$, which requires the workload to supply enough columns to occupy them.
At $M{=}1$ a $32\times32$ core flattens to $1\times1024$, and a device instantiating sixteen such cores needs more still.
Many decoding GEMMs fall short; we call a shape with $M < m$ and $N < n$ a \emph{narrow and flat} GEMM.
Here no selector pattern helps: shrinking $n$ to match $N$ leaves the rows beyond the first few idle, while keeping $n$ large leaves their columns empty.
What remains is $K$, which stays large in exactly these cases.
Mapping part of that reduction onto the array is therefore the only way to recover utilization, and it is the third dimension of the MEU introduced in Section~\ref{subsec:array}.

\subsection{Partitioning the Reduction}

The array is divided into $t$ groups along $K$, each assigned a contiguous slice of the reduction and each producing an independent partial sum for the \emph{same} output tile.
The PE budget is then shared three ways,
\begin{equation}
m \cdot n \cdot t = S^2, \qquad k = t \cdot \max(m,n),
\label{eq:splitk}
\end{equation}
so the temporal extent grows with the number of groups: the core consumes $t$ slices of depth $\max(m,n)$ before every PE has contributed.
Figure~\ref{fig:splitk} shows $M{=}2$, $N{=}4$ on a core that would otherwise flatten to $2\times8$.
Since $N{=}4$ cannot fill eight columns, the configuration uses $m{=}2$, $n{=}4$, and $t{=}2$ instead, giving $\mathbf{MEU = 2\times4\times8}$ at the same PE count.

Physically nothing new is required for the compute phase.
Each group is an ordinary $m \times n$ sub-array executing an ordinary output-stationary GEMM over its own slice of $K/t$, driven by the same selector pattern used for spatial regrouping.
The weight buffer supplies $t$ streams exactly as it would for $t$ spatial groups, so the delivery path is unchanged.

\subsection{Tick-Tock Reduction}

Splitting along $K$ differs from splitting along $M$ or $N$ in one respect that shapes the entire design.
Output tiles produced by different $m$ or $n$ groups are independent and simply drain in parallel.
Partial sums produced by different $k$ groups are not: they are addends of the same result and must be reduced before leaving the array.

A dedicated adder tree is the obvious solution and an impractical one.
Reducing across $t$ groups spanning the array height requires a tree for every one of the $n$ columns, together with wiring that gathers operands from PEs at arbitrary vertical distances.
At array scale this approaches the cost of a second PE array, and its routing is precisely the long-distance, many-to-one interconnect that dominates physical design.

\dyna{} instead reduces along the vertical path the array already has.
The final group is held \emph{stationary} and serves as the destination accumulator, while every other group releases its partial sums downward one row per cycle.
A \emph{tick} moves a partial sum one row down without accumulating; a \emph{tock} adds the arriving value into the local accumulator, which happens once it reaches its counterpart in the group below.
A logical row of width $n$ occupies $n/S$ physical rows and a group holds $m$ of them, so one group spans
\begin{equation}
h \;=\; \frac{m \cdot n}{S} \;=\; \frac{S}{t}
\label{eq:groupheight}
\end{equation}
physical rows, using $m \cdot n \cdot t = S^2$ from Equation~\ref{eq:splitk}.
In Figure~\ref{fig:splitk}, $t{=}2$ on a $4\times4$ core gives $h{=}2$: one tick and one tock.

\subsection{Cost at the Output Path}

The scheme is inexpensive because it is the array's existing drain, with one change.
In the baseline design, every PE holds a result and the array empties row by row, each value passing downward through the PEs beneath it, which takes $S$ cycles.
Under Split-$K$ the same $S$ cycles are spent, but the values passing through the final group are added into it rather than forwarded past it.
Reduction and evacuation are thus the same phase: the array still clears in $S$ cycles, and what leaves it is the reduced result rather than $t$ separate partial sums.

%% file: content/5.prefill.tex
\section{Disaggregated Quantization}
\label{sec:quantization}

Prefill is compute-bound and benefits from dual-side quantization (W8A8), while decoding is memory-bound and benefits from weight-only quantization (W4A16).
Serving both on one substrate requires PEs that execute either without duplicating hardware.
This section shows that the way this is done matters as much as the fact that it is done: the conventional construction inflates the output path, whereas ours leaves it untouched and, in doing so, matches both the asymmetric regrouping of Section~\ref{sec:array} and the Split-$K$ mapping of Section~\ref{sec:splitk}.

\begin{figure}[t]
    \centering
    \includegraphics[width=0.95\linewidth]{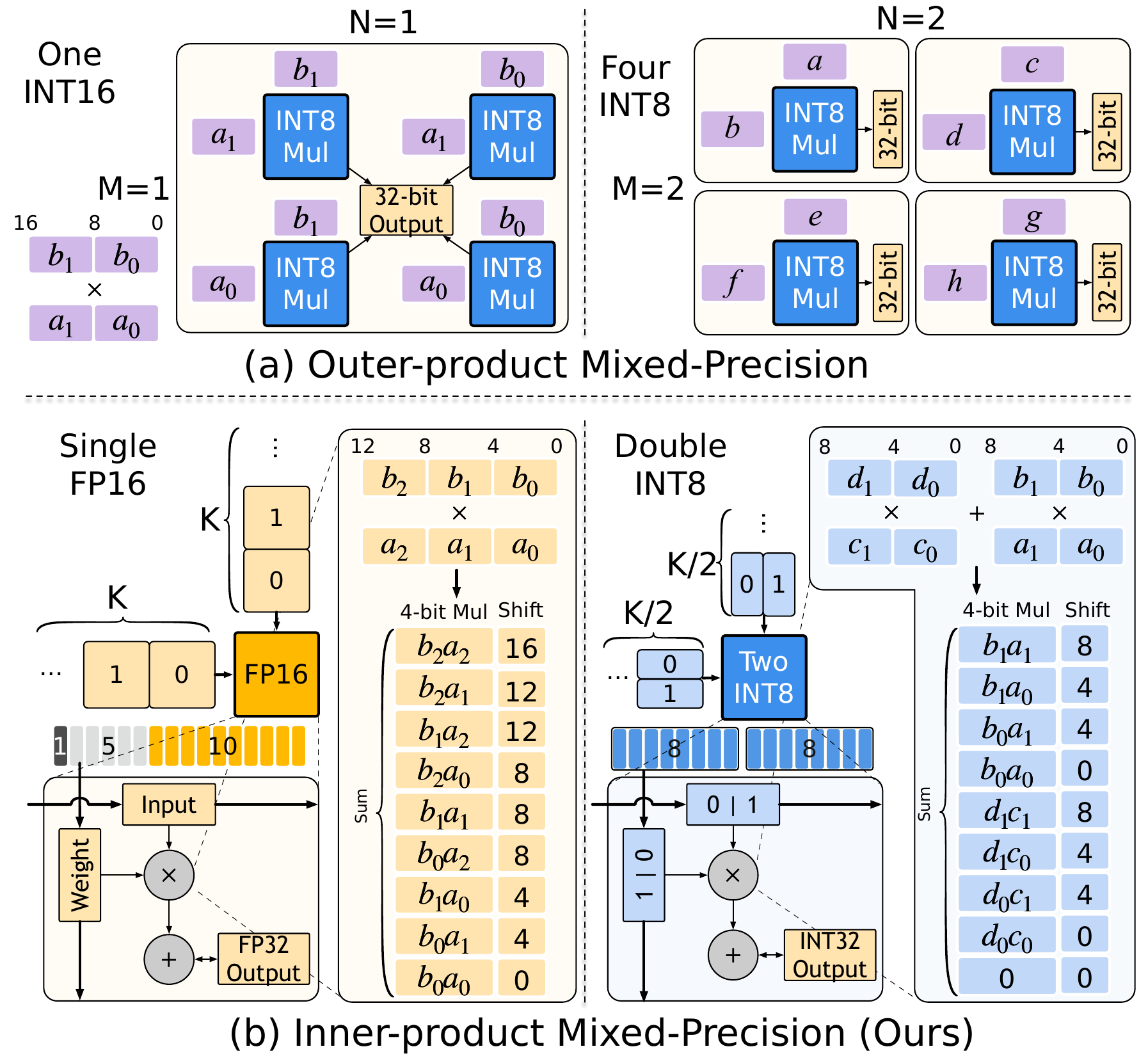}
    \vspace{-2mm}
\caption{
Mixed-precision construction.
(a) Outer-product schemes expand the output tile from $1{\times}1$ to $2{\times}2$.
(b) Ours packs operands along $K$, leaving the output tile at $1{\times}1$ in both modes.
}
    \vspace{-5mm}
    \label{fig:mixed}
\end{figure}

\subsection{Outer-Product versus Inner-Product Construction}

Bit-composable PEs~\cite{sharma2018bitfusion,guo2022ant} build a high-precision multiply from low-precision sub-multipliers, and at reduced precision re-task those sub-multipliers to carry more work.
The question is where that work goes.

\textbf{Outer product.}
In the conventional choice (Figure~\ref{fig:mixed}(a)), one INT16 multiply decomposes into the four cross terms of $\{a_1,a_0\} \otimes \{b_1,b_0\}$, shifted and summed into a single output; at INT8 the same four multipliers are re-interpreted as four \emph{unrelated} operand pairs that share no reduction.
Throughput quadruples, but the output tile expands from $1{\times}1$ to $2{\times}2$, so the PE needs four accumulators instead of one.
Our synthesis results show that this construction costs over $2\times$ the area of a plain INT16 unit at equal multiplier count, since the added accumulators and output registers, not the multipliers, dominate.
The penalty also compounds with the floating-point path: with four independent adders already committed to separate outputs, none can be folded into an FP16 accumulation, so FP16 support would require a datapath of its own rather than reusing what is there.
Together, these costs make the construction impractical.

\textbf{Inner product.}
\dyna{} instead assigns the freed sub-multipliers to operands \emph{adjacent along $K$}, which the GEMM already reduces.
An FP16 multiply needs an 11-bit mantissa product, implemented as a 12-bit multiplier built from nine 4-bit sub-multipliers combined by shift-and-add; two INT8 multiplies need eight of those same nine (Figure~\ref{fig:mixed}(b)).
The INT8 mode therefore reuses the FP16 structure almost entirely, packing a pair of $K$-adjacent operands into the 16-bit input word and summing their products in place, exactly as the output-stationary dataflow would have summed them across two cycles.
The PE still holds one accumulator and emits one 32-bit result; only the effective depth consumed per cycle changes with precision.
Operand widths at the array boundary are unchanged, so nothing about the mapping, the drain, or the interconnect is affected.

\subsection{Matching the Array Design}

\textbf{Asymmetric regrouping.}
Flattening the array multiplies the weight bandwidth required while leaving the input path at a single broadcast stream, and weight-only quantization relieves exactly that side: at 4 bits the same buffer port supplies four times as many elements per cycle, while activations stay in FP16 on the path that needed no extra bandwidth.
Keeping activations at FP16 precision also confines dequantization to the buffer interface, so no dequantization logic enters the PEs even when all $S^2$ of them are fed independently.

\textbf{Split-$K$ mapping.}
Split-$K$ and the inner-product datapath make the same choice at two levels: the former spends surplus PEs on the reduction dimension, the latter spends surplus multiplier capacity there.
Both land where the GEMM already reduces, so the work is absorbed by existing accumulators rather than by new outputs.
Precision therefore acts only on $k$: packing two INT8 operands per cycle doubles the reduction depth a PE covers, leaving $m$, $n$, and $t$ and the reduction schedule untouched.

\textbf{Summary.}
A single datapath supports W8A8 prefill and W4A16 decoding by directing low-precision throughput along $K$ rather than into additional outputs.
Output width and count are invariant to precision; weight bandwidth falls by $4\times$ where the flattened configuration needs it most, and neither the array shape, the drain path, nor the reduction schedule is affected.

%% file: content/5.5.scheduling.tex
\section{System--Architecture Co-Scheduling}
\label{sec:scheduling}

\begin{figure}[t]
    \centering
    \includegraphics[width=0.95\linewidth]{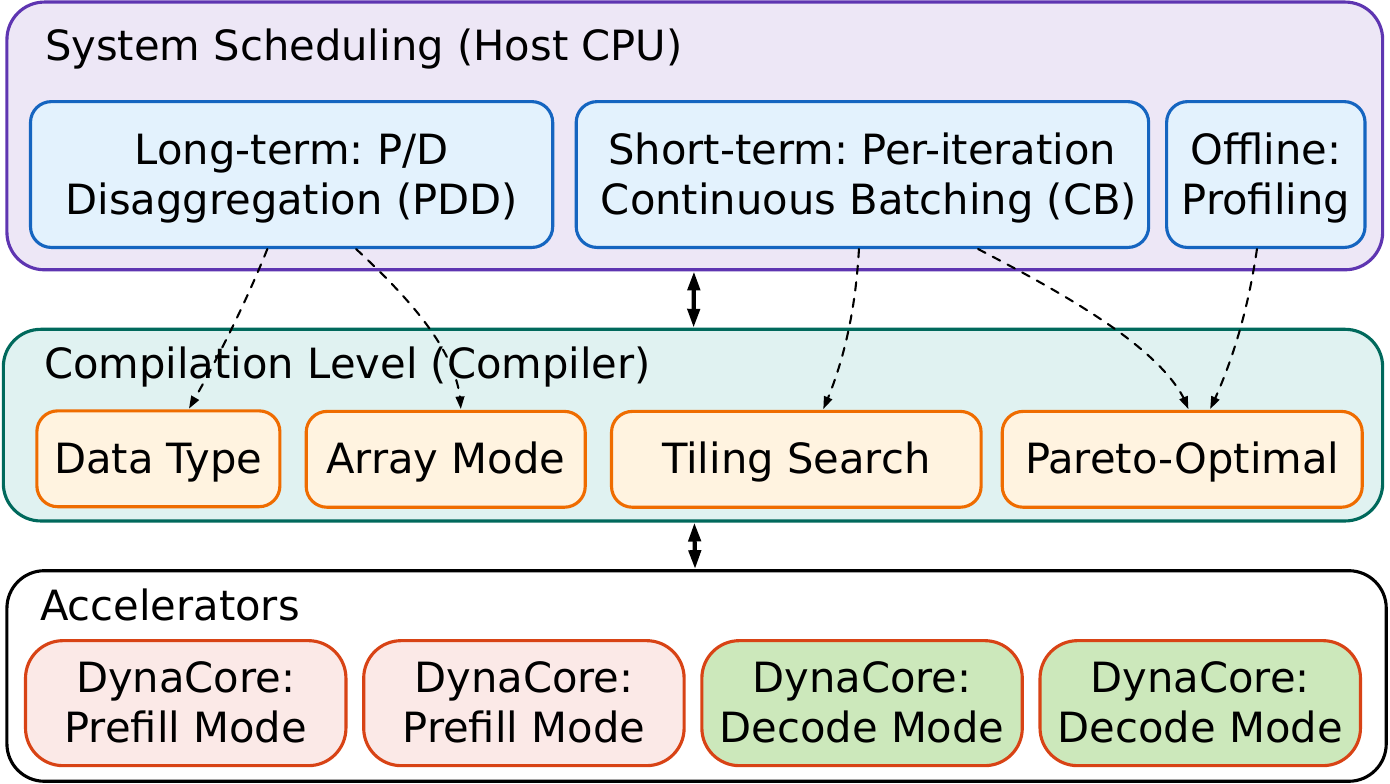}
    \vspace{-2mm}
    \caption{\dyna{} three-layer co-scheduling. The host scheduler handles long-term PDD and short-term CB; the compiler selects an MEU configuration from offline profiling; the accelerators execute under the selected configuration.}
    \label{fig:sched}
    \vspace{-6mm}
\end{figure}

The reshaping mechanisms of Sections~\ref{sec:array} and~\ref{sec:splitk} expose a configuration space that must be navigated at runtime.
\dyna{} does so through three layers: a host \emph{system scheduler}, a \emph{compilation} layer that selects an MEU for each batch, and the \emph{accelerators}.
Figure~\ref{fig:sched} gives an overview.
Prior work has studied system-level scheduling extensively, so we omit low-level serving heuristics and focus on the interaction between scheduling and the proposed architecture.

\subsection{System Scheduling}
\label{sec:sched-system}

The top layer runs on the host CPU and supports continuous batching (CB) and prefill--decoding disaggregation (PDD) over multiple devices through abstract interfaces to the backend accelerators.
It operates at two time scales.

\textbf{Long-term scheduling.}
PDD assigns each device a phase, and the optimal prefill/decoding device ratio varies with the workload: long-input traces favor prefill capacity, long-output traces favor decoding.
\dyna{} adapts this ratio from workload statistics and fixes the per-phase quantization setting accordingly, which we evaluate in Section~\ref{sec:multi}.

\textbf{Short-term scheduling.}
Within each iteration, CB determines which requests are batched together, and thus the token count that reaches the accelerator.
This count is what the compilation layer consumes: it changes from iteration to iteration, and each value implies a different MEU.

\subsection{Compilation Layer}
\label{sec:sched-compile}

Both control paths feed a compilation layer that maps the assigned batch onto concrete MEUs.
Long-term scheduling fixes the data type; the compiler then chooses how to decompose the batch and which extents $(m,n,t)$ to use for each part.

\begin{wrapfigure}{b}{0.55\linewidth}
    \centering
    \includegraphics[width=\linewidth]{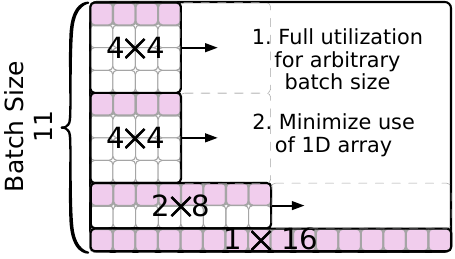}
    \vspace{-4mm}
    \caption{Pyramid packing: $M{=}11$ split into four MEUs on a $4{\times}4$ core.}
    \vspace{-2mm}
    \label{fig:Pyramid}
\end{wrapfigure}

\textbf{Pyramid packing.}
A batch size need not map onto a single MEU.
When $M$ falls between two reachable configurations, forcing one shape either idles rows or splits the batch across passes, so \dyna{} instead decomposes the batch into several MEUs executed within the same scheduling window.
Figure~\ref{fig:Pyramid} shows $M{=}11$ on a $4{\times}4$ core decomposed into $4{\times}4$, $4{\times}4$, $2{\times}8$, and $1{\times}16$, filling the array exactly.
The decomposition is hierarchical, taking the largest configuration that fits at each step, which keeps utilization high at arbitrary batch sizes while using the fully flattened configuration only for the remainder.
The latter matters because flattening trades weight reuse for occupancy, so minimizing its share reduces memory traffic and energy.

\textbf{Configuration search.}
Packing fixes $m$ for each part; the remaining freedom is how to divide the rest of the PE budget between $n$ and $t$.
Several choices reach comparable utilization while differing in memory behavior: a larger $n$ covers more output columns per pass but consumes weight bandwidth, whereas a larger $t$ maps more of the reduction onto the array and suits shapes whose $N$ is bounded by the model.
Total energy follows the same tension, since the wider configuration draws higher instantaneous power yet can finish sooner.
The compiler therefore scores every candidate satisfying $m \cdot n \cdot t = S^2$ against a joint latency--energy objective; \Alg{alg:tiling} states the procedure.

\begin{algorithm2e}[t]
\caption{Batch decomposition and MEU selection}
\label{alg:tiling}
\KwIn{shape $(M,N,K)$; core dimension $S$; energy weight $\lambda$; profiling tables $\mathcal{T}_{\mathrm{lat}}$, $\mathcal{T}_{\mathrm{eng}}$}
\KwOut{list $\mathcal{P}$ of split configurations $(m,n,t)$}
$\mathcal{P} \leftarrow \emptyset$; \quad $r \leftarrow M$\;
\While{$r > 0$}{
  $m \leftarrow \max\{\,2^i \le \min(r,S)\,\}$ \tcp*{split $M$}
  $c^\star \leftarrow \infty$\;
  \ForEach{$t \in \{1,2,4,\ldots,S^2/m\}$}{
    $n \leftarrow S^2/(m\,t)$ \tcp*{split $N$}
    $k \leftarrow t \cdot \max(m,n)$ \tcp*{split $K$}
    $\tau \leftarrow \lceil N/n \rceil \cdot \lceil K/k \rceil$ \tcp*{passes}
    $c \leftarrow \tau \cdot \big( \mathcal{T}_{\mathrm{lat}}[m,n,t] + \lambda\, \mathcal{T}_{\mathrm{eng}}[m,n,t] \big)$\;
    \If{$c < c^\star$}{
      $c^\star \leftarrow c$; \quad $(n^\star,t^\star) \leftarrow (n,t)$\;
    }
  }
  $\mathcal{P} \leftarrow \mathcal{P} \cup \{(m,n^\star,t^\star)\}$; \quad $r \leftarrow r - m$\;
}
\Return $\mathcal{P}$\;
\end{algorithm2e}

\textbf{Offline profiling.}
Both tables are populated offline, once per core design, by measuring each reachable configuration on representative shapes; $\lambda$ sets the latency--energy preference and is fixed per deployment.
The search space is small, since $m$, $n$, and $t$ are powers of two constrained by $m \cdot n \cdot t = S^2$, so a $32{\times}32$ core admits only tens of candidates and the compiler enumerates them exhaustively.
At runtime the result is a table lookup keyed by batch size rather than a search, which keeps the cost negligible relative to an iteration.

%% file: content/6.evaluation.tex
\section{Evaluation}

\subsection{Methodology}

\textbf{Models and Datasets.}
We evaluate \dyna{} on several modern LLMs, including Llama2-7B, Llama2-13B~\cite{touvron2023llama2}, and Llama3.1-8B~\cite{llama2024llama3}. We benchmark these models on ShareGPT~\cite{sharegpt2023} and Dolly~\cite{DatabricksBlog2023DollyV2} datasets, as well as production traces from MoonCake~\cite{qin2024mooncake} (a conversation trace and a tool interaction trace), and a synthetic dataset with fixed input/output lengths. These datasets collectively represent diverse LLM inference scenarios with varying prompt lengths and generation lengths, covering both short-form and long-form conversational workloads.
We adopt GPTQ~\cite{frantar2022gptq} for disaggregated quantization, a widely used LLM quantization method, to quantize the model to W8A8 for prefill and W4A16 for decoding.

\textbf{Baseline Architectures.}
We compare \dyna{} against several representative GEMM architectures for DNN and LLM acceleration. These include BitFusion (BF)~\cite{sharma2018bitfusion}, a systolic architecture for low-precision DNNs, FIGLUT~\cite{park2025figlut}, an accelerator that leverages look-up tables for FP-INT GEMM in quantized LLMs, and Planaria~\cite{ghodrati2020planaria} and SARA~\cite{samajdar2022self}, reconfigurable systolic-array accelerators for DNN and LLM inference. In addition, we evaluate a systolic array with our proposed disaggregated quantization (DQ) for comparison.
The detailed configurations of all architectures are summarized in the top part of \Tbl{tab:baselines}.

\textbf{System and Architecture Implementation.}
We build a full-stack simulation framework that models both the frontend system behavior and the backend hardware execution. The frontend simulates an online LLM serving environment, supporting continuous batching and prefill–decoding disaggregation, enabling accurate microsecond-level serving dynamics. The backend implements a cycle-accurate simulator for \dyna{} and all baseline architectures, capturing compute behavior and memory access under real LLM workloads.

We implement the \dyna{} architecture in SystemVerilog and synthesize it using Synopsys Design Compiler with the TSMC 28~nm standard-cell library to obtain area overhead, leakage power, and dynamic power.
The on-chip buffers are evaluated using CACTI~7.0~\cite{rajeev2017cacti7}, while off-chip memory behavior and energy are modeled with DRAMSim3~\cite{li2020dramsim3} and integrated into our cycle-accurate simulator. The overall architectural setup is summarized in Table~\ref{tab:setup}.
For a fair comparison, all baseline accelerators are evaluated with the same methodology.

\begin{table}[t]
\centering
\caption{Comparison of baseline architectures and \dyna{}. ``Same'' indicates the same configuration as prefill. Baselines use identical precision across prefill and decoding, while \dyna{} adopts phase-specific configurations. All designs use 28nm, 500MHz, an 8MB buffer, and 4TB/s of DRAM bandwidth.}
\vspace{-1mm}
\resizebox{\linewidth}{!}{
\begin{tabular}{l|cc|ccc|cc}
\toprule
\multirow{2}{*}{\textbf{Architecture}} & \multicolumn{2}{c|}{\textbf{PE Array}} & \multicolumn{3}{c|}{\textbf{Precision}} & \textbf{Area} & \textbf{Power} \\
& Shape & Unit & Prefill & Decode & KV & (mm$^2$) & (W) \\
\midrule
BitFusion~\cite{sharma2018bitfusion} & $128{\times}128$ & INT8 MAC & W8A8 & Same & KV8-Q8 & 23.64 & 12.89 \\
SA~\cite{kung1979systolic} w/ \textit{DQ} & $128{\times}128$ & \textit{DQ} MAC & W8A8 & W4A16 & KV4-Q16 & 65.09 & 111.55 \\
FIGLUT~\cite{park2025figlut} & $4{\times}(16{\times}16)$ & 16b LUT & W4A16 & Same & KV4-Q16 & 42.36 & 89.26 \\
Planaria~\cite{ghodrati2020planaria} & $128{\times}128$ & INT8 MAC & W8A8 & Same & KV8-Q8 & 70.98 & 133.37 \\
SARA~\cite{samajdar2022self} & $128{\times}128$ & INT8 MAC & W8A8 & Same & KV8-Q8 & 68.83 & 164.49 \\
\midrule
\textbf{\dyna{}} & $16{\times}(32{\times}32)$ & \textit{DQ} MAC & W8A8 & W4A16 & KV4-Q16 & 76.44 & 115.74 \\
\bottomrule
\end{tabular}
}
\vspace{-3mm}
\label{tab:baselines}
\end{table}

\begin{table}[t]
    \centering
    \caption{\dyna Setup}
    \vspace{-3mm}
    \resizebox{\linewidth}{!}{
    \begin{tabular}{l|c}
        \toprule
        PE Array & $16 \times (32 \times 32)$; Output Stationary \\
        \midrule
        Vector Unit & $16 \times $ (EXP $\times32$; MUL $\times32$; ADD $\times32$; SQRT $\times1$; DIV $\times1$)\\
        \midrule
        \multirow{2}{*}{Quantization} & Prefill: Input INT8, Weight INT8, KV INT8 \\
                     & Decoding: Input INT8/FP16, Weight 4-bit, KV 4-bit \\
        \midrule
        Buffer & Input $16\times$128KB/32Banks; Weight/Output $16\times$192KB/32Banks \\
        \midrule
        Memory & HBM4E: 4.0 TB/s, 2,048-bit interface, 16 Gb/s/pin \\
        \bottomrule
    \end{tabular}
    }
    \vspace{-3mm}
    \label{tab:setup}
\end{table}

\begin{figure*}[t]
    \centering
    \includegraphics[width=1.0\textwidth]{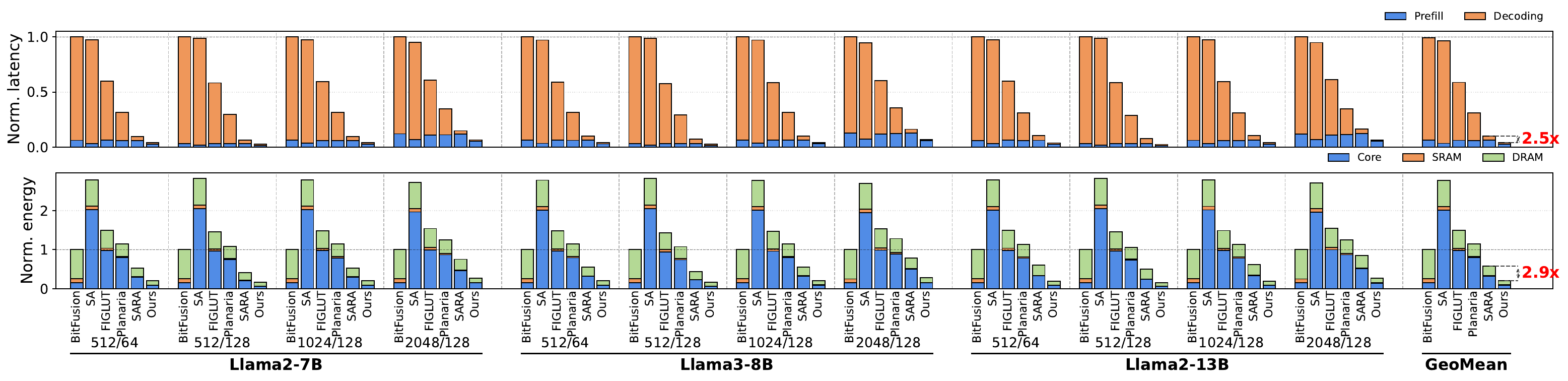}
    \vspace{-5mm}
    \caption{
    End-to-end latency and energy consumption comparison (lower is better).
    The numbers XX/YY shown below each group indicate the input and output lengths, respectively: XX denotes the prompt length, while YY denotes the generated output length.
}
    \vspace{-10pt}
    \label{fig:perf}
\end{figure*}

\subsection{LLM Accuracy Comparison}

In \Tbl{tab:accuracy}, we evaluate our proposed \emph{disaggregated quantization} method on the Llama3.1-8B-Instruct model. We compare our approach with weight-only quantization algorithm, GPTQ~\cite{frantar2022gptq} (W4A16), as well as the dual-side quantization algorithms, Qserve~\cite{lin2024qserve} (W4A8) and GPTQ~\cite{frantar2022gptq} (W8A8). Additionally, we include BCQ~\cite{park2025unifying}, which is supported by FIGLUT~\cite{park2025figlut}. Our evaluations cover several downstream datasets, including ARC-Easy and ARC-Challenge~\cite{clark2018think}, BoolQ~\cite{clark2019boolq}, HellaSwag~\cite{zellers2019hellaswag}, PIQA~\cite{bisk2020piqa}, and GSM8K~\cite{cobbe2021training}.

Overall, our \emph{disaggregated quantization} method, which utilizes W8A8 for prefill and W4A16 for decoding, incurs an average accuracy drop of only \textbf{0.07\%} compared to FP16. This drop is significantly less than that of GPTQ (W4A16), which shows a \textbf{2.81\%} drop, and QServe (W4A8), which demonstrates a drop of \textbf{1.51\%}. Our method's accuracy drop is close to that of GPTQ (W8A8), which is \textbf{0.09\%}. These results illustrate that our approach achieves a balance between accuracy and efficiency, preserving model quality while attaining high performance in both the prefill and decoding phases.

\begin{table}[t]
\centering
\caption{Accuracy comparison on LLaMA3.1-8B-Instruct across downstream tasks.}
\vspace{-2mm}
\resizebox{\linewidth}{!}{
\begin{tabular}{lcccccc}
\toprule
\textbf{Method} & \textbf{ARC-C} & \textbf{ARC-E} & \textbf{BoolQ} & \textbf{HellaSwag} & \textbf{PIQA} & \textbf{GSM8K} \\
\midrule
FP16              & 55.03 & 79.63 & 84.13 & 79.27 & 80.96 & 75.66 \\
W8A8-GPTQ         & 55.12 & 79.63 & 83.91 & \textbf{79.05} & 80.41 & 76.04 \\
W4A16-GPTQ        & 51.62 & 75.93 & 82.87 & 77.78 & 79.11 & 70.51 \\
W4A8-QServe       & 53.41 & 79.29 & 83.58 & 77.99 & 79.65 & 71.72 \\
W4A16-FIGLUT      & --    & --    & --    & --    & --    & 72.38 \\
\midrule
\textbf{Ours}     & \textbf{55.03} & \textbf{79.76} & \textbf{84.01} & 78.32 & \textbf{81.01} & \textbf{76.12} \\
\bottomrule
\end{tabular}
}
\label{tab:accuracy}
\vspace{-6mm}
\end{table}

\subsection{Architecture Performance Comparison}

We evaluate \dyna{} and all baseline architectures under a single-device setup to assess their standalone efficiency on LLM inference. Each workload consists of one input sequence with a typical input prompt and output generation length. We measure the end-to-end latency and energy normalized by BitFusion. The overall results are shown in \Fig{fig:perf}.

\textbf{End-to-End Latency.}
As shown in the top of \Fig{fig:perf}, \dyna{} achieves the lowest end-to-end latency among all baselines. Compared with the systolic array with DQ, \dyna{} delivers a $23.07\times$ speedup, primarily due to the decoding phase. By selecting the optimal hardware configuration from compilation, \dyna{} maintains high utilization even at batch size 1, whereas the systolic array's fixed $128\times128\times128$ MEU leads to substantial underutilization.

Compared with BitFusion and FIGLUT, \dyna{} gains a $23.81\times$ and $14.13\times$ speedup, benefiting from the adaptive dataflow and higher INT8 throughput.
We further compare against the reconfigurable systolic arrays Planaria and SARA, over which \dyna{} achieves $7.57\times$ and $2.48\times$ end-to-end speedup. Although both reshape their arrays to improve decoding utilization, their coarse reconfiguration granularity ($32\times32$ fission for Planaria and $4\times4$ cells for SARA) still leaves the array underutilized on batch-1 GEMV. With finer-grained reconfiguration together with split-$K$, \dyna{} provides a $23.03\times$ and $3.48\times$ decoding-phase speedup over Planaria and SARA, respectively.

\textbf{Energy.} The bottom of \Fig{fig:perf} shows the energy breakdown for all architectures. \dyna{} achieves the best energy efficiency, outperforming BitFusion, Systolic Array, FIGLUT, Planaria, and SARA by $4.96\times$, $13.76\times$, $7.39\times$, $5.71\times$, and $2.86\times$, respectively. These gains mainly follow from the latency reductions discussed above. Although \dyna{} has slightly higher on-chip power due to extra hardware support for splittable array and DQ, its shorter execution time reduces leakage energy and yields the highest overall efficiency.

\textbf{Hardware Overhead.}
The area and on-chip power of \dyna{} and all baselines are summarized in the last two columns of \Tbl{tab:baselines}.
Compared to BitFusion INT8, the systolic array with \textit{DQ} incurs higher area and power because its processing elements support both INT8 and FP16.
\dyna{} introduces further overhead over the systolic \textit{DQ} design due to its shape-adaptive execution.
Overall, these mechanisms introduce moderate overhead that is well justified by the latency improvements reported above.

\subsection{Batching Analysis}

\begin{figure}[t]
    \centering
    \includegraphics[width=0.48\textwidth]{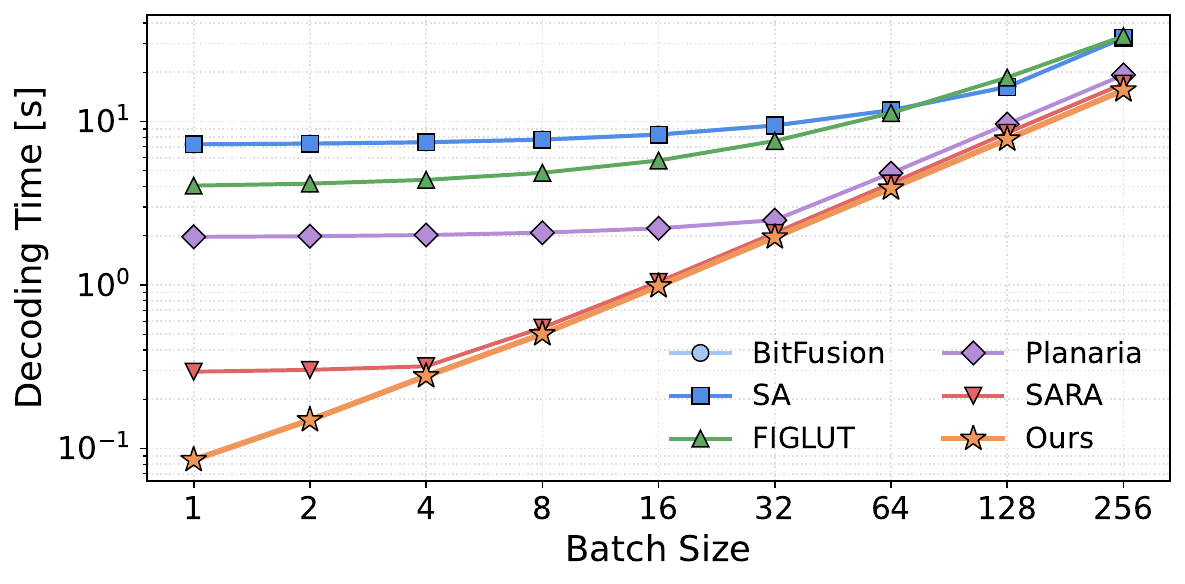}
    \vspace{-3mm}
    \caption{Decoding Latency under differing batch sizes}
    \vspace{-5mm}
    \label{fig:batch}
\end{figure}

To further characterize \dyna{} across different regimes, we evaluate the decoding latency of Llama-3-8B on a single device with batch sizes ranging from $1$ to $256$. As shown in \Fig{fig:batch}, latency remains nearly constant while the PE array is underutilized and increases approximately linearly once the array becomes compute-bound.

The fixed $128\times128$ systolic array reaches saturation at a batch size of approximately $128$. The reconfigurable designs saturate earlier: Planaria at around $32$, SARA at around $4$, and \dyna{} at around $1$. \dyna{} reaches high utilization even at small batch sizes because its fine-grained array partitioning and split-$K$ execution expose substantially more parallelism. Beyond saturation, \dyna{} retains an advantage over the baselines because its compiler selects the most efficient configuration and scheduling strategy for each workload regime.

Consequently, \dyna{} achieves the lowest latency across the entire batch-size range, reducing latency by $85.2\times$ over the systolic array at batch size $1$ and by $2.09\times$ at batch size $256$. Even at large batch size, it achieves the lowest $60.7\,\mathrm{ms}$ per-request latency, compared with $66.5\,\mathrm{ms}$ for SARA, $75.2\,\mathrm{ms}$ for Planaria, and $126.9\,\mathrm{ms}$ for the systolic array.

\subsection{Multi-Device Evaluation for LLM Serving}
\label{sec:multi}
Beyond single-device performance, we evaluate \dyna{} in a multi-device online serving environment and focus on service-level objective (SLO) metrics that directly reflect user-perceived latency.

\textbf{Evaluation Setup.}
We simulate an 8-device system handling incoming LLM requests.
Each device is a chip instance of a given architecture with off-chip HBM, and devices are interconnected with 16~GB/s links, modeled following~\cite{alian2018simulating}; this inter-device communication cost is incurred whenever prefill and decoding are partitioned across separate devices.
Unlike the single-device setting where prefill and decoding run on the same chip, here we follow the prefill--decoding disaggregation paradigm~\cite{patel2024splitwise,zhong2024distserve}, which improves SLO attainment.
Several devices are assigned to prefill, and the remaining devices process decoding tasks.

The system is evaluated using the real-world datasets running Llama3.1-8B.
For datasets without arrival timestamps, such as ShareGPT and Dolly, request arrivals follow a Poisson process.
For MoonCake Conversation and MoonCake Tool, we use the real production timestamps.
The scheduler uses a first-come-first-serve policy: prefill requests form batches according to arrival time, while decoding requests use continuous batching to improve utilization.

We measure two SLO metrics: Time to First Token (TTFT), which is mainly determined by prefill latency, and Time Per Output Token (TPOT), which reflects decoding performance.

\begin{figure}[t]
\centering
\includegraphics[width=0.45\textwidth]{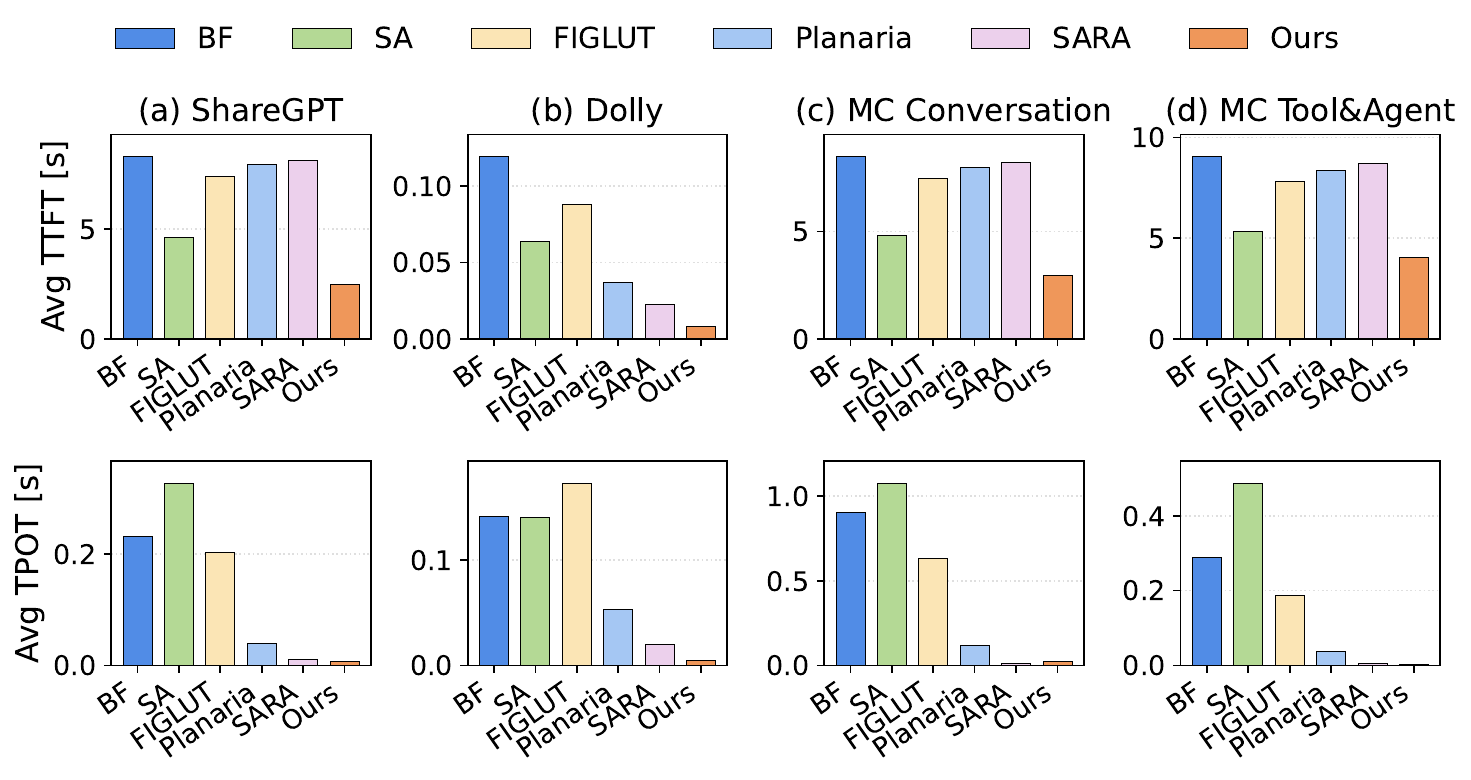}
\vspace{-3mm}
\caption{Average TTFT and TPOT on real-world datasets. The Y-Axis values are in seconds.}
% \vspace{-3mm}
\label{fig:ttft}
\end{figure}

\textbf{System-level Comparison.}
We evaluate homogeneous multi-device systems in \Fig{fig:ttft}, using 4 prefill and 4 decoding devices for ShareGPT and Dolly, and 6 prefill and 2 decoding devices for MoonCake.

\dyna{} achieves the lowest TTFT across all workloads. Baselines such as Planaria and FIGLUT remain prefill-bound because they do not accelerate large prefill GEMMs, whereas \dyna{} combines disaggregated quantization with shape-adaptive execution. This yields a $2.97\times$ and $3.50\times$ average TTFT improvement over Planaria and FIGLUT, respectively.

\dyna{} also achieves the best overall TPOT. Reconfigurable arrays such as Planaria are the strongest baselines, while FIGLUT is over an order of magnitude slower on average. For TPOT, \dyna{} achieves an $8.02\times$ and $36.55\times$ average improvement over Planaria and FIGLUT, respectively. Its fine-grained splitting, Split-$K$, and pyramid packing maintain high utilization for the small decoding batches common in online serving, delivering substantial SLO improvements across realistic workloads.

\begin{figure}[t]
\centering
\includegraphics[width=0.48\textwidth]{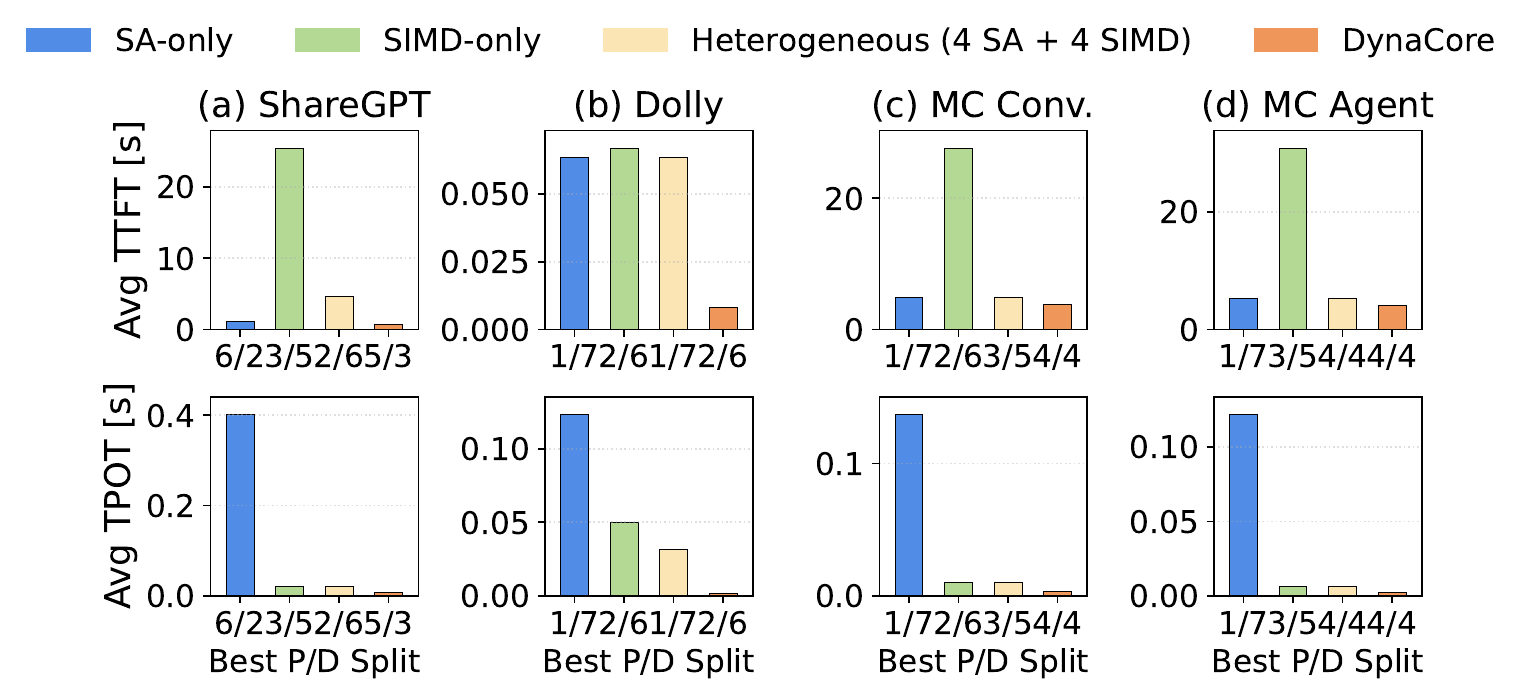}
\vspace{-3mm}
\caption{Comparison of \dyna{} with static homogeneous and heterogeneous multi-device systems.}

% \vspace{-6mm}
\label{fig:hetero}
\end{figure}

\textbf{\dyna{} compared with static system.} \Fig{fig:hetero} compares \dyna{} against three static eight-device systems: an SA-only design optimized for large-$M$ GEMMs, a SIMD-only design optimized for small-$M$ GEMMs and GEMVs, and a heterogeneous system comprising four SA devices and four SIMD devices. Here, SIMD denotes a vector-style architecture that applies the same instruction across multiple lanes and is particularly effective for GEMV-dominated decoding. Each system contains the same total number of PEs as \dyna{}.

The heterogeneous system serves as a strong baseline because it includes devices specialized for both prefill and decoding. However, its fixed 4:4 hardware partition becomes suboptimal when the ideal allocation between prefill and decoding devices varies across workloads. For example, on the decode-heavy Dolly workload, the heterogeneous design achieves a TPOT of $0.032\,\mathrm{s}$, compared with only $0.0017\,\mathrm{s}$ for \dyna{}. In contrast, \dyna{} dynamically reconfigures the same silicon resources at runtime, enabling it to match or outperform all static configurations in both TTFT and TPOT across all workload traces.

\subsection{Ablation Study}

\begin{figure}[t]
\centering
\includegraphics[width=0.45\textwidth]{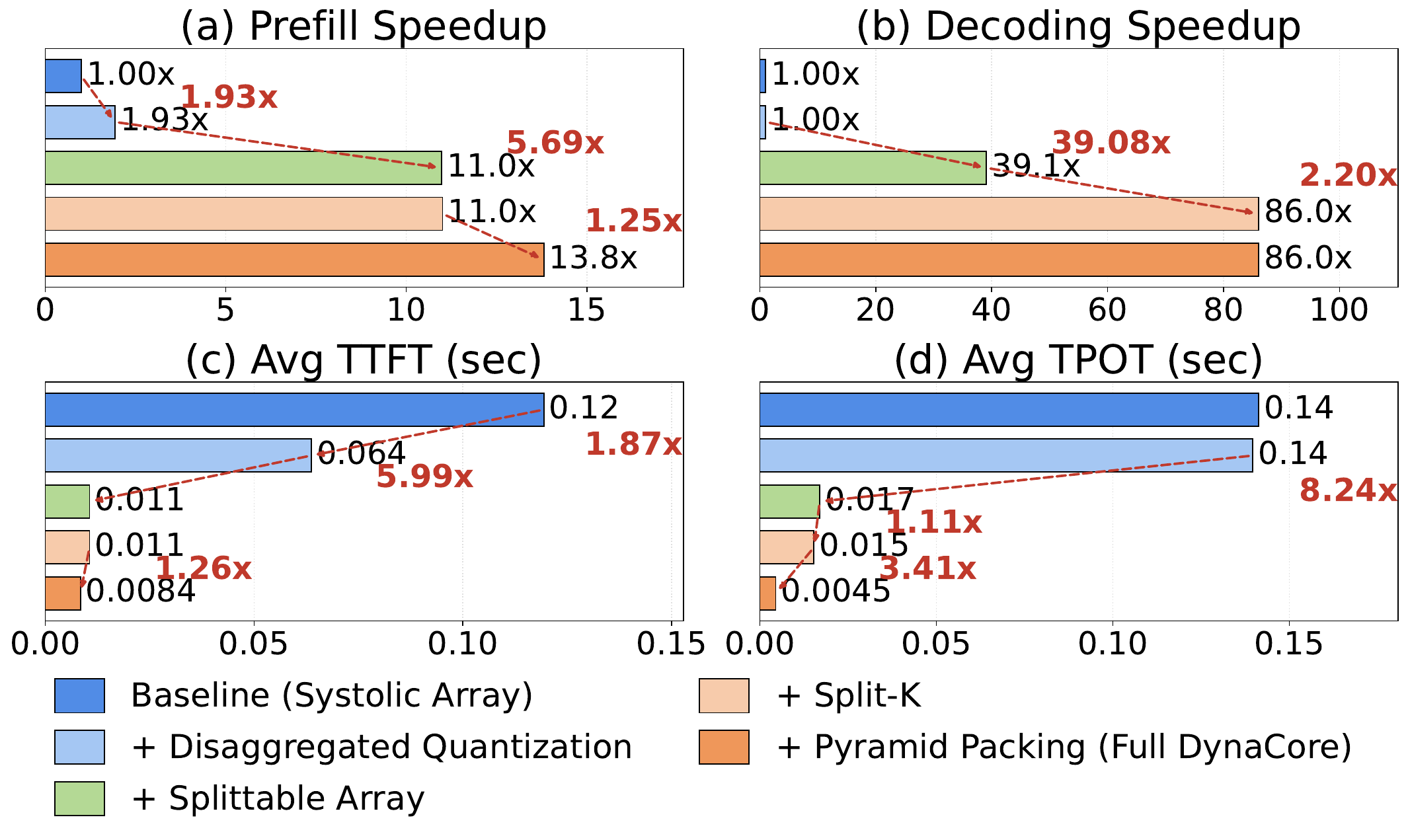}
% \vspace{-4mm}
\caption{Ablation study of \dyna
}
\vspace{-5mm}
\label{fig:ablation}
\end{figure}

We progressively add each \dyna{} component to a vanilla $128{\times}128$ systolic array running Llama2-7B, and evaluate performance on the Dolly dataset, as shown in \Fig{fig:ablation}.
Because the components are added cumulatively, each bar reports a component's marginal gain given the components before it, and a component's benefit can surface only once an earlier bottleneck is removed.

\textbf{Prefill.}
For single-device inference at batch size 1 (\Fig{fig:ablation}(a)), disaggregated quantization, the splittable array, and pyramid packing improve prefill by $1.93\times$, $5.69\times$, and $1.25\times$, respectively.
The $1.93\times$ from DQ reflects the doubled INT8 throughput of W8A8.
The large splittable-array gain stems from the short prompts of Dolly, which average only 17.9 tokens: with $M$ far below the array dimension, prefill GEMMs are flat rather than tall, and the monolithic array suffers the same row underutilization as in decoding.
For the long prompts in \Fig{fig:perf} (512--2048 tokens), prefill GEMMs are tall and the square configuration suffices, so the end-to-end gains there come almost entirely from decoding, consistent with Section~\ref{sec:array}.

\textbf{Decoding.}
In decoding (\Fig{fig:ablation}(b)), DQ alone yields no speedup despite cutting weight traffic by $4\times$.
On the vanilla array, a batch-1 GEMV activates one row of PEs, and the resulting compute time exceeds even the FP16 weight-streaming time by about $31\times$ (Section~\ref{subsec:quant}); latency is bound by utilization, not bandwidth, so reducing traffic has no effect.
The splittable array removes this bottleneck and delivers a $39.1\times$ speedup, which in turn exposes the benefit of DQ: with FP16 weights, the bandwidth floor would cap the splittable-array gain at about $31\times$, so the excess is attributable to the 4-bit weights.
Split-$K$ adds another $2.20\times$ by improving utilization for small GEMM shapes.

\textbf{Online serving.}
In multi-device online serving (\Fig{fig:ablation}(c)(d)), the same components improve TTFT by $1.87\times$, $5.99\times$, and $1.26\times$, respectively.
For TPOT, DQ again yields no gain on the vanilla array for the same reason, while the splittable array, Split-$K$, and pyramid packing provide improvements of $8.24\times$, $1.11\times$, and $3.41\times$.
Pyramid packing matters far more here than at batch size 1, since continuous batching produces irregular $M$ that a single MEU cannot tile efficiently.

%% file: content/7.conclusion.tex
\section{Conclusion}

We present \dyna{}, a phase-adaptive architecture for efficient LLM serving. By co-designing array organization, quantization, and runtime scheduling, \dyna{} aligns hardware execution with the distinct characteristics of prefill and decoding.
Our design integrates a splittable systolic array, disaggregated quantization, and adaptive scheduling into a unified framework. Evaluation shows that \dyna{} substantially reduces service-level latency over FIGLUT and Planaria, improving TTFT by up to 3.50$\times$ and 2.97$\times$ and TPOT by up to 36.55$\times$ and 8.02$\times$, respectively, under real workloads.
These results demonstrate the effectiveness of phase-aware system--architecture co-design for next-generation LLM serving.

%% file: refs.bib
@article{hugo2023llama,
  author       = {Hugo Touvron and
                  Thibaut Lavril and
                  Gautier Izacard and
                  Xavier Martinet and
                  Marie{-}Anne Lachaux and
                  Timoth{\'{e}}e Lacroix and
                  Baptiste Rozi{\`{e}}re and
                  Naman Goyal and
                  Eric Hambro and
                  Faisal Azhar and
                  Aur{\'{e}}lien Rodriguez and
                  Armand Joulin and
                  Edouard Grave and
                  Guillaume Lample},
  title        = {LLaMA: Open and Efficient Foundation Language Models},
  journal      = {CoRR},
  volume       = {abs/2302.13971},
  year         = {2023},
  url          = {https://doi.org/10.48550/arXiv.2302.13971},
  doi          = {10.48550/ARXIV.2302.13971},
  eprinttype    = {arXiv},
  eprint       = {2302.13971},
  bibsource    = {dblp computer science bibliography, https://dblp.org}
}

@inproceedings{park2025unifying,
  title={Unifying uniform and binary-coding quantization for accurate compression of large language models},
  author={Park, Seungcheol and Bae, Jeongin and Kwon, Beomseok and Kim, Minjun and Kim, Byeongwook and Kwon, Se Jung and Kang, U and Lee, Dongsoo},
  booktitle={Proceedings of the 63rd Annual Meeting of the Association for Computational Linguistics (Volume 1: Long Papers)},
  pages={28468--28488},
  year={2025}
}

@article{touvron2023llama2,
  author    = {Touvron, Hugo and Martin, Louis and Stone, Kevin and Albert, Peter and Almahairi, Amjad and Babaei, Yasmine and Bashlykov, Nikolay and Batra, Soumya and Bhargava, Prajjwal and Bhosale, Shruti and Bikel, Dan and Blecher, Lukas and Ferrer, Cristian Canton and Chen, Moya and Cucurull, Guillem and Esiobu, David and Fernandes, Jude and Fu, Jeremy and Fu, Wenyin and Fuller, Brian and Gao, Cynthia and Goswami, Vedanuj and Goyal, Naman and Hartshorn, Anthony and Hosseini, Saghar and Hou, Rui and Inan, Hakan and Kardas, Marcin and Kerkez, Viktor and Khabsa, Madian and Kloumann, Isabel and Korenev, Artem and Koura, Punit Singh and Lachaux, Marie-Anne and Lavril, Thibaut and Lee, Jenya and Liskovich, Diana and Lu, Yinghai and Mao, Yuning and Martinet, Xavier and Mihaylov, Todor and Mishra, Pushkar and Molybog, Igor and Nie, Yixin and Poulton, Andrew and Reizenstein, Jeremy and Rungta, Rashi and Saladi, Kalyan and Schelten, Alan and Silva, Ruan and Smith, Eric Michael and Subramanian, Ranjan and Tan, Xiaoqing Ellen and Tang, Binh and Taylor, Ross and Williams, Adina and Kuan, Jian Xiang and Xu, Puxin and Yan, Zheng and Zarov, Iliyan and Zhang, Yuchen and Fan, Angela and Kambadur, Melanie and Narang, Sharan and Rodriguez, Aurelien and Stojnic, Robert and Edunov, Sergey and Scialom, Thomas},
  title        = {Llama 2: Open Foundation and Fine-Tuned Chat Models},
  journal      = {CoRR},
  volume       = {abs/2307.09288},
  year         = {2023},
  url          = {https://doi.org/10.48550/arXiv.2307.09288},
  doi          = {10.48550/ARXIV.2307.09288},
  eprinttype    = {arXiv},
  eprint       = {2307.09288},
  bibsource    = {dblp computer science bibliography, https://dblp.org}
}

@article{radford2019language_gpt2,
  title={Language Models are Unsupervised Multitask Learners},
  author={Radford, Alec and Wu, Jeff and Child, Rewon and Luan, David and Amodei, Dario and Sutskever, Ilya},
  year={2019}
}

@article{jiang2023mistral_7b,
  author       = {Albert Q. Jiang and
                  Alexandre Sablayrolles and
                  Arthur Mensch and
                  Chris Bamford and
                  Devendra Singh Chaplot and
                  Diego de Las Casas and
                  Florian Bressand and
                  Gianna Lengyel and
                  Guillaume Lample and
                  Lucile Saulnier and
                  L{\'{e}}lio Renard Lavaud and
                  Marie{-}Anne Lachaux and
                  Pierre Stock and
                  Teven Le Scao and
                  Thibaut Lavril and
                  Thomas Wang and
                  Timoth{\'{e}}e Lacroix and
                  William El Sayed},
  title        = {Mistral 7B},
  journal      = {CoRR},
  volume       = {abs/2310.06825},
  year         = {2023},
  url          = {https://doi.org/10.48550/arXiv.2310.06825},
  doi          = {10.48550/ARXIV.2310.06825},
  eprinttype    = {arXiv},
  eprint       = {2310.06825},
}

@inproceedings{vaswani2017attention,
  author       = {Ashish Vaswani and
                  Noam Shazeer and
                  Niki Parmar and
                  Jakob Uszkoreit and
                  Llion Jones and
                  Aidan N. Gomez and
                  Lukasz Kaiser and
                  Illia Polosukhin},
  editor       = {Isabelle Guyon and
                  Ulrike von Luxburg and
                  Samy Bengio and
                  Hanna M. Wallach and
                  Rob Fergus and
                  S. V. N. Vishwanathan and
                  Roman Garnett},
  title        = {Attention is All you Need},
  booktitle    = {Advances in Neural Information Processing Systems 30: Annual Conference
                  on Neural Information Processing Systems 2017, December 4-9, 2017,
                  Long Beach, CA, {USA}},
  pages        = {5998--6008},
  year         = {2017},
  url          = {https://proceedings.neurips.cc/paper/2017/hash/3f5ee243547dee91fbd053c1c4a845aa-Abstract.html},
}

@inproceedings{kwon2023efficient_vllm,
  title={Efficient Memory Management for Large Language Model Serving with PagedAttention},
  author={Woosuk Kwon and Zhuohan Li and Siyuan Zhuang and Ying Sheng and Lianmin Zheng and Cody Hao Yu and Joseph E. Gonzalez and Hao Zhang and Ion Stoica},
  booktitle={Proceedings of the ACM SIGOPS 29th Symposium on Operating Systems Principles},
  year={2023}
}

@inproceedings{zheng2024sglang,
author = {Zheng, Lianmin and Yin, Liangsheng and Xie, Zhiqiang and Sun, Chuyue and Huang, Jeff and Yu, Cody Hao and Cao, Shiyi and Kozyrakis, Christos and Stoica, Ion and Gonzalez, Joseph E. and Barrett, Clark and Sheng, Ying},
title = {SGLang: efficient execution of structured language model programs},
year = {2024},
isbn = {9798331314385},
publisher = {Curran Associates Inc.},
address = {Red Hook, NY, USA},
booktitle = {Proceedings of the 38th International Conference on Neural Information Processing Systems},
articleno = {2000},
numpages = {27},
location = {Vancouver, BC, Canada},
series = {NIPS '24}
}

@inproceedings{guo2022ant,
  author       = {Cong Guo and
                  Chen Zhang and
                  Jingwen Leng and
                  Zihan Liu and
                  Fan Yang and
                  Yunxin Liu and
                  Minyi Guo and
                  Yuhao Zhu},
  title        = {{ANT:} Exploiting Adaptive Numerical Data Type for Low-bit Deep Neural
                  Network Quantization},
  booktitle    = {55th {IEEE/ACM} International Symposium on Microarchitecture, {MICRO}
                  2022, Chicago, IL, USA, October 1-5, 2022},
  pages        = {1414--1433},
  publisher    = {{IEEE}},
  year         = {2022},
  url          = {https://doi.org/10.1109/MICRO56248.2022.00095},
  doi          = {10.1109/MICRO56248.2022.00095},
  bibsource    = {dblp computer science bibliography, https://dblp.org}
}

@inproceedings{guo2023olive,
  title={Olive: Accelerating large language models via hardware-friendly outlier-victim pair quantization},
  author={Guo, Cong and Tang, Jiaming and Hu, Weiming and Leng, Jingwen and Zhang, Chen and Yang, Fan and Liu, Yunxin and Guo, Minyi and Zhu, Yuhao},
  booktitle={Proceedings of the 50th Annual International Symposium on Computer Architecture},
  pages={1--15},
  year={2023}
}

@inproceedings{guo2025transitive,
  title={Transitive Array: An Efficient GEMM Accelerator with Result Reuse},
  author={Guo, Cong and Wei, Chiyue and Tang, Jiaming and Duan, Bowen and Han, Song and Li, Hai and Chen, Yiran},
  booktitle={Proceedings of the 52nd Annual International Symposium on Computer Architecture},
  pages={990--1004},
  year={2025}
}

@INPROCEEDINGS{jang2024figna,
  author={Jang, Jaeyong and Kim, Yulhwa and Lee, Juheun and Kim, Jae-Joon},
  booktitle={2024 IEEE International Symposium on High-Performance Computer Architecture (HPCA)}, 
  title={FIGNA: Integer Unit-Based Accelerator Design for FP-INT GEMM Preserving Numerical Accuracy}, 
  year={2024},
  volume={},
  number={},
  pages={760-773},
  doi={10.1109/HPCA57654.2024.00064}
}

@INPROCEEDINGS{park2025figlut,
  author={Park, Gunho and Kwon, Hyeokjun and Kim, Jiwoo and Bae, Jeongin and Park, Baeseong and Lee, Dongsoo and Lee, Youngjoo},
  booktitle={2025 IEEE International Symposium on High Performance Computer Architecture (HPCA)}, 
  title={FIGLUT: An Energy-Efficient Accelerator Design for FP-INT GEMM Using Look-Up Tables}, 
  year={2025},
  volume={},
  number={},
  pages={1098-1111},
  doi={10.1109/HPCA61900.2025.00085}}

@INPROCEEDINGS{hu2025mant,
  author={Hu, Weiming and Zhang, Haoyan and Guo, Cong and Feng, Yu and Guan, Renyang and Hua, Zhendong and Liu, Zihan and Guan, Yue and Guo, Minyi and Leng, Jingwen},
  booktitle={2025 IEEE International Symposium on High Performance Computer Architecture (HPCA)}, 
  title={M-ANT: Efficient Low-bit Group Quantization for LLMs via Mathematically Adaptive Numerical Type}, 
  year={2025},
  volume={},
  number={},
  pages={1112-1126},
  doi={10.1109/HPCA61900.2025.00086}}

@inproceedings{chen2025bitmod,
  author       = {Yuzong Chen and
                  Ahmed F. AbouElhamayed and
                  Xilai Dai and
                  Yang Wang and
                  Marta Andronic and
                  George A. Constantinides and
                  Mohamed S. Abdelfattah},
  title        = {BitMoD: Bit-serial Mixture-of-Datatype {LLM} Acceleration},
  booktitle    = {{IEEE} International Symposium on High Performance Computer Architecture,
                  {HPCA} 2025, Las Vegas, NV, USA, March 1-5, 2025},
  pages        = {1082--1097},
  publisher    = {{IEEE}},
  year         = {2025},
  url          = {https://doi.org/10.1109/HPCA61900.2025.00084},
  doi          = {10.1109/HPCA61900.2025.00084},
  bibsource    = {dblp computer science bibliography, https://dblp.org}
}

@inproceedings{yu2022orca,
  author       = {Gyeong{-}In Yu and
                  Joo Seong Jeong and
                  Geon{-}Woo Kim and
                  Soojeong Kim and
                  Byung{-}Gon Chun},
  editor       = {Marcos K. Aguilera and
                  Hakim Weatherspoon},
  title        = {Orca: {A} Distributed Serving System for Transformer-Based Generative
                  Models},
  booktitle    = {16th {USENIX} Symposium on Operating Systems Design and Implementation,
                  {OSDI} 2022, Carlsbad, CA, USA, July 11-13, 2022},
  pages        = {521--538},
  publisher    = {{USENIX} Association},
  year         = {2022},
  url          = {https://www.usenix.org/conference/osdi22/presentation/yu},
  bibsource    = {dblp computer science bibliography, https://dblp.org}
}

@inproceedings{zhong2024distserve,
  author       = {Yinmin Zhong and
                  Shengyu Liu and
                  Junda Chen and
                  Jianbo Hu and
                  Yibo Zhu and
                  Xuanzhe Liu and
                  Xin Jin and
                  Hao Zhang},
  editor       = {Ada Gavrilovska and
                  Douglas B. Terry},
  title        = {DistServe: Disaggregating Prefill and Decoding for Goodput-optimized
                  Large Language Model Serving},
  booktitle    = {18th {USENIX} Symposium on Operating Systems Design and Implementation,
                  {OSDI} 2024, Santa Clara, CA, USA, July 10-12, 2024},
  pages        = {193--210},
  publisher    = {{USENIX} Association},
  year         = {2024},
  url          = {https://www.usenix.org/conference/osdi24/presentation/zhong-yinmin},
  bibsource    = {dblp computer science bibliography, https://dblp.org}
}

@inproceedings{patel2024splitwise,
  author       = {Pratyush Patel and
                  Esha Choukse and
                  Chaojie Zhang and
                  Aashaka Shah and
                  {\'{I}}{\~{n}}igo Goiri and
                  Saeed Maleki and
                  Ricardo Bianchini},
  title        = {Splitwise: Efficient Generative {LLM} Inference Using Phase Splitting},
  booktitle    = {51st {ACM/IEEE} Annual International Symposium on Computer Architecture,
                  {ISCA} 2024, Buenos Aires, Argentina, June 29 - July 3, 2024},
  pages        = {118--132},
  publisher    = {{IEEE}},
  year         = {2024},
  url          = {https://doi.org/10.1109/ISCA59077.2024.00019},
  doi          = {10.1109/ISCA59077.2024.00019},
  bibsource    = {dblp computer science bibliography, https://dblp.org}
}

@article{frantar2022gptq,
  author       = {Elias Frantar and
                  Saleh Ashkboos and
                  Torsten Hoefler and
                  Dan Alistarh},
  title        = {{GPTQ:} Accurate Post-Training Quantization for Generative Pre-trained
                  Transformers},
  journal      = {CoRR},
  volume       = {abs/2210.17323},
  year         = {2022},
  url          = {https://doi.org/10.48550/arXiv.2210.17323},
  doi          = {10.48550/ARXIV.2210.17323},
  eprinttype    = {arXiv},
  eprint       = {2210.17323},
  bibsource    = {dblp computer science bibliography, https://dblp.org}
}

@inproceedings{xiao2023smoothquant,
  author       = {Guangxuan Xiao and
                  Ji Lin and
                  Micka{\"{e}}l Seznec and
                  Hao Wu and
                  Julien Demouth and
                  Song Han},
  editor       = {Andreas Krause and
                  Emma Brunskill and
                  Kyunghyun Cho and
                  Barbara Engelhardt and
                  Sivan Sabato and
                  Jonathan Scarlett},
  title        = {SmoothQuant: Accurate and Efficient Post-Training Quantization for
                  Large Language Models},
  booktitle    = {International Conference on Machine Learning, {ICML} 2023, 23-29 July
                  2023, Honolulu, Hawaii, {USA}},
  series       = {Proceedings of Machine Learning Research},
  volume       = {202},
  pages        = {38087--38099},
  publisher    = {{PMLR}},
  year         = {2023},
  url          = {https://proceedings.mlr.press/v202/xiao23c.html},
  bibsource    = {dblp computer science bibliography, https://dblp.org}
}

@article{lin2024awq,
  title={Awq: Activation-aware weight quantization for on-device llm compression and acceleration},
  author={Lin, Ji and Tang, Jiaming and Tang, Haotian and Yang, Shang and Chen, Wei-Ming and Wang, Wei-Chen and Xiao, Guangxuan and Dang, Xingyu and Gan, Chuang and Han, Song},
  journal={Proceedings of machine learning and systems},
  volume={6},
  pages={87--100},
  year={2024}
}

@article{dettmers2022gpt3,
  title={Gpt3. int8 (): 8-bit matrix multiplication for transformers at scale},
  author={Dettmers, Tim and Lewis, Mike and Belkada, Younes and Zettlemoyer, Luke},
  journal={Advances in neural information processing systems},
  volume={35},
  pages={30318--30332},
  year={2022}
}

@article{lin2024qserve,
  title={Qserve: W4a8kv4 quantization and system co-design for efficient llm serving},
  author={Lin, Yujun and Tang, Haotian and Yang, Shang and Zhang, Zhekai and Xiao, Guangxuan and Gan, Chuang and Han, Song},
  journal={arXiv preprint arXiv:2405.04532},
  year={2024}
}

@inproceedings{chen2024bbs,
  title={BBS: Bi-directional bit-level sparsity for deep learning acceleration},
  author={Chen, Yuzong and Meng, Jian and Seo, Jae-sun and Abdelfattah, Mohamed S},
  booktitle={2024 57th IEEE/ACM International Symposium on Microarchitecture (MICRO)},
  pages={551--564},
  year={2024},
  organization={IEEE}
}

@article{rajeev2017cacti7,
  author       = {Rajeev Balasubramonian and
                  Andrew B. Kahng and
                  Naveen Muralimanohar and
                  Ali Shafiee and
                  Vaishnav Srinivas},
  title        = {{CACTI} 7: New Tools for Interconnect Exploration in Innovative Off-Chip
                  Memories},
  journal      = {{ACM} Trans. Archit. Code Optim.},
  volume       = {14},
  number       = {2},
  pages        = {14:1--14:25},
  year         = {2017},
  url          = {https://doi.org/10.1145/3085572},
  doi          = {10.1145/3085572},
  bibsource    = {dblp computer science bibliography, https://dblp.org}
}

@article{li2020dramsim3,
  author       = {Shang Li and
                  Zhiyuan Yang and
                  Dhiraj Reddy and
                  Ankur Srivastava and
                  Bruce L. Jacob},
  title        = {DRAMsim3: {A} Cycle-Accurate, Thermal-Capable {DRAM} Simulator},
  journal      = {{IEEE} Comput. Archit. Lett.},
  volume       = {19},
  number       = {2},
  pages        = {110--113},
  year         = {2020},
  url          = {https://doi.org/10.1109/LCA.2020.2973991},
  doi          = {10.1109/LCA.2020.2973991},
  bibsource    = {dblp computer science bibliography, https://dblp.org}
}

@inproceedings{sharma2018bitfusion,
  author       = {Hardik Sharma and
                  Jongse Park and
                  Naveen Suda and
                  Liangzhen Lai and
                  Benson Chau and
                  Vikas Chandra and
                  Hadi Esmaeilzadeh},
  editor       = {Murali Annavaram and
                  Timothy Mark Pinkston and
                  Babak Falsafi},
  title        = {Bit Fusion: Bit-Level Dynamically Composable Architecture for Accelerating
                  Deep Neural Network},
  booktitle    = {45th {ACM/IEEE} Annual International Symposium on Computer Architecture,
                  {ISCA} 2018, Los Angeles, CA, USA, June 1-6, 2018},
  pages        = {764--775},
  publisher    = {{IEEE} Computer Society},
  year         = {2018},
  url          = {https://doi.org/10.1109/ISCA.2018.00069},
  doi          = {10.1109/ISCA.2018.00069},
  bibsource    = {dblp computer science bibliography, https://dblp.org}
}

@article{llama2024llama3,
  author       = {Llama Team},
  title        = {The Llama 3 Herd of Models},
  journal      = {CoRR},
  volume       = {abs/2407.21783},
  year         = {2024},
  url          = {https://doi.org/10.48550/arXiv.2407.21783},
  doi          = {10.48550/ARXIV.2407.21783},
  eprinttype    = {arXiv},
  eprint       = {2407.21783},
  bibsource    = {dblp computer science bibliography, https://dblp.org}
}

@article{Devlin2018BERT,
  author       = {Jacob Devlin and
                  Ming{-}Wei Chang and
                  Kenton Lee and
                  Kristina Toutanova},
  title        = {{BERT:} Pre-training of Deep Bidirectional Transformers for Language
                  Understanding},
  journal      = {CoRR},
  volume       = {abs/1810.04805},
  year         = {2018},
  url          = {http://arxiv.org/abs/1810.04805},
  eprinttype    = {arXiv},
  eprint       = {1810.04805},
  bibsource    = {dblp computer science bibliography, https://dblp.org}
}

@online{DatabricksBlog2023DollyV2,
    author    = {Mike Conover and Matt Hayes and Ankit Mathur and Jianwei Xie and Jun Wan and Sam Shah and Ali Ghodsi and Patrick Wendell and Matei Zaharia and Reynold Xin},
    title     = {Free Dolly: Introducing the World's First Truly Open Instruction-Tuned LLM},
    year      = {2023},
    url       = {https://www.databricks.com/blog/2023/04/12/dolly-first-open-commercially-viable-instruction-tuned-llm},
    urldate   = {2023-06-30}
}

@article{samajdar2018scale,
  title={SCALE-Sim: Systolic CNN Accelerator Simulator},
  author={Samajdar, Ananda and Zhu, Yuhao and Whatmough, Paul and Mattina, Matthew and Krishna, Tushar},
  journal={arXiv preprint arXiv:1811.02883},
  year={2018}
}

@misc{sharegpt2023,
  author       = {ShareGPT Team},
  title        = {ShareGPT},
  howpublished = {\url{https://sharegpt.com/}},
  year         = {2023}
}

@article{thakur2024verigen,
author = {Thakur, Shailja and Ahmad, Baleegh and Pearce, Hammond and Tan, Benjamin and Dolan-Gavitt, Brendan and Karri, Ramesh and Garg, Siddharth},
title = {VeriGen: A Large Language Model for Verilog Code Generation},
year = {2024},
issue_date = {May 2024},
publisher = {Association for Computing Machinery},
address = {New York, NY, USA},
volume = {29},
number = {3},
issn = {1084-4309},
url = {https://doi.org/10.1145/3643681},
doi = {10.1145/3643681},
journal = {ACM Trans. Des. Autom. Electron. Syst.},
month = apr,
articleno = {46},
numpages = {31}
}

@article{deepseek2025deepseek_r1,
  author       = {DeepSeek{-}AI},
  title        = {DeepSeek-R1: Incentivizing Reasoning Capability in LLMs via Reinforcement
                  Learning},
  journal      = {CoRR},
  volume       = {abs/2501.12948},
  year         = {2025},
  url          = {https://doi.org/10.48550/arXiv.2501.12948},
  doi          = {10.48550/ARXIV.2501.12948},
  eprinttype    = {arXiv},
  eprint       = {2501.12948},
  bibsource    = {dblp computer science bibliography, https://dblp.org}
}

@inproceedings{xiong2025hlspilot,
author = {Xiong, Chenwei and Liu, Cheng and Li, Huawei and Li, Xiaowei},
title = {HLSPilot: LLM-based High-Level Synthesis},
year = {2025},
isbn = {9798400710773},
publisher = {Association for Computing Machinery},
address = {New York, NY, USA},
url = {https://doi.org/10.1145/3676536.3676781},
doi = {10.1145/3676536.3676781},
booktitle = {Proceedings of the 43rd IEEE/ACM International Conference on Computer-Aided Design},
articleno = {226},
numpages = {9},
location = {Newark Liberty International Airport Marriott, New York, NY, USA},
series = {ICCAD '24}
}

@article{zhuo2024bigcodebench,
  title={BigCodeBench: Benchmarking Code Generation with Diverse Function Calls and Complex Instructions},
  author    = {Zhuo, Terry Yue and Vu, Minh Chien and Chim, Jenny and Hu, Han and Yu, Wenhao and Widyasari, Ratnadira and Yusuf, Imam Nur Bani and Zhan, Haolan and He, Junda and Paul, Indraneil and Brunner, Simon and Gong, Chen and Hoang, Thong and Zebaze, Armel Randy and Hong, Xiaoheng and Li, Wen-Ding and Kaddour, Jean and Xu, Ming and Zhang, Zhihan and Yadav, Prateek and Jain, Naman and Gu, Alex and Cheng, Zhoujun and Liu, Jiawei and Liu, Qian and Wang, Zijian and Hui, Binyuan and Muennighoff, Niklas and Lo, David and Fried, Daniel and Du, Xiaoning and de Vries, Harm and Von Werra, Leandro},
  journal={arXiv preprint arXiv:2406.15877},
  year={2024}
}

@misc{OpenAI2023GPT4,
  author       = {OpenAI},
  title        = {GPT-4 Technical Report},
  year         = {2023},
  howpublished = {\url{https://cdn.openai.com/papers/gpt-4.pdf}},
  note         = {Accessed: 2025-11-16}
}

@article{chen2025dpad,
  author       = {Xinhua Chen and
                  Sitao Huang and
                  Cong Guo and
                  Chiyue Wei and
                  Yintao He and
                  Jianyi Zhang and
                  Hai Helen Li and
                  Yiran Chen},
  title        = {DPad: Efficient Diffusion Language Models with Suffix Dropout},
  journal      = {CoRR},
  volume       = {abs/2508.14148},
  year         = {2025},
  url          = {https://doi.org/10.48550/arXiv.2508.14148},
  doi          = {10.48550/ARXIV.2508.14148},
  eprinttype    = {arXiv},
  eprint       = {2508.14148},
  bibsource    = {dblp computer science bibliography, https://dblp.org}
}

@inproceedings{katharopoulos2020transformers_are_rnns,
  author       = {Angelos Katharopoulos and
                  Apoorv Vyas and
                  Nikolaos Pappas and
                  Fran{\c{c}}ois Fleuret},
  title        = {Transformers are RNNs: Fast Autoregressive Transformers with Linear
                  Attention},
  booktitle    = {Proceedings of the 37th International Conference on Machine Learning,
                  {ICML} 2020, 13-18 July 2020, Virtual Event},
  series       = {Proceedings of Machine Learning Research},
  volume       = {119},
  pages        = {5156--5165},
  publisher    = {{PMLR}},
  year         = {2020},
  url          = {http://proceedings.mlr.press/v119/katharopoulos20a.html},
  bibsource    = {dblp computer science bibliography, https://dblp.org}
}

@inproceedings{zhu2024multilingual,
  author       = {Wenhao Zhu and
                  Hongyi Liu and
                  Qingxiu Dong and
                  Jingjing Xu and
                  Shujian Huang and
                  Lingpeng Kong and
                  Jiajun Chen and
                  Lei Li},
  editor       = {Kevin Duh and
                  Helena G{\'{o}}mez{-}Adorno and
                  Steven Bethard},
  title        = {Multilingual Machine Translation with Large Language Models: Empirical
                  Results and Analysis},
  booktitle    = {Findings of the Association for Computational Linguistics: {NAACL}
                  2024, Mexico City, Mexico, June 16-21, 2024},
  pages        = {2765--2781},
  publisher    = {Association for Computational Linguistics},
  year         = {2024},
  url          = {https://doi.org/10.18653/v1/2024.findings-naacl.176},
  doi          = {10.18653/V1/2024.FINDINGS-NAACL.176},
  bibsource    = {dblp computer science bibliography, https://dblp.org}
}

@inproceedings{samajdar2022self,
  title={Self adaptive reconfigurable arrays (sara) learning flexible gemm accelerator configuration and mapping-space using ml},
  author={Samajdar, Ananda and Qin, Eric and Pellauer, Michael and Krishna, Tushar},
  booktitle={Proceedings of the 59th ACM/IEEE Design Automation Conference},
  pages={583--588},
  year={2022}
}

@article{wang2020linformer,
  author       = {Sinong Wang and
                  Belinda Z. Li and
                  Madian Khabsa and
                  Han Fang and
                  Hao Ma},
  title        = {Linformer: Self-Attention with Linear Complexity},
  journal      = {CoRR},
  volume       = {abs/2006.04768},
  year         = {2020},
  url          = {https://arxiv.org/abs/2006.04768},
  eprinttype    = {arXiv},
  eprint       = {2006.04768},
  bibsource    = {dblp computer science bibliography, https://dblp.org}
}

@article{holmes2024deepspeed,
  title={Deepspeed-fastgen: High-throughput text generation for llms via mii and deepspeed-inference},
  author    = {Holmes, Connor and Tanaka, Masahiro and Wyatt, Michael and Awan, Ammar Ahmad and Rasley, Jeff and Rajbhandari, Samyam and Aminabadi, Reza Yazdani and Qin, Heyang and Bakhtiari, Arash and Kurilenko, Lev and He, Yuxiong},
  journal={arXiv preprint arXiv:2401.08671},
  year={2024}
}

@inproceedings{alian2018simulating,
  title={Simulating PCI-Express interconnect for future system exploration},
  author={Alian, Mohammad and Srinivasan, Krishna Parasuram and Kim, Nam Sung},
  booktitle={2018 IEEE International Symposium on Workload Characterization (IISWC)},
  pages={168--178},
  year={2018},
  organization={IEEE}
}

@inproceedings{kung1979systolic,
  title={Systolic arrays (for VLSI)},
  author={Kung, Hsiang Tsung and Leiserson, Charles E},
  booktitle={Sparse Matrix Proceedings 1978},
  volume={1},
  pages={256--282},
  year={1979},
  organization={Society for industrial and applied mathematics Philadelphia, PA, USA}
}

@article{radford2018improving,
  title={Improving language understanding by generative pre-training},
  author    = {Radford, Alec and Narasimhan, Karthik and Salimans, Tim and Sutskever, Ilya},
  year={2018},
  publisher={San Francisco, CA, USA}
}

@article{achiam2023gpt,
  title={Gpt-4 technical report},
  author    = {Achiam, Josh and Adler, Steven and Agarwal, Sandhini and Ahmad, Lama and Akkaya, Ilge and Aleman, Florencia Leoni and Almeida, Diogo and Altenschmidt, Janko and Altman, Sam and Anadkat, Shyamal and Avila, Red and Babuschkin, Igor and Balaji, Suchir and Balcom, Valerie and Baltescu, Paul and Bao, Haiming and Bavarian, Mohammad and Belgum, Jeff and Bello, Irwan and Berdine, Jake and Bernadett-Shapiro, Gabriel and Berner, Christopher and Bogdonoff, Lenny and Boiko, Oleg and Boyd, Madelaine and Brakman, Anna-Luisa and Brockman, Greg and Brooks, Tim and Brundage, Miles and Button, Kevin and Cai, Trevor and Campbell, Rosie and Cann, Andrew and Carey, Brittany and Carlson, Chelsea and Carmichael, Rory and Chan, Brooke and Chang, Che and Chantzis, Fotis and Chen, Derek and Chen, Sully and Chen, Ruby and Chen, Jason and Chen, Mark and Chess, Ben and Cho, Chester and Chu, Casey and Chung, Hyung Won and Cummings, Dave and Currier, Jeremiah and Dai, Yunxing and Decareaux, Cory and Degry, Thomas and Deutsch, Noah and Deville, Damien and Dhar, Arka and Dohan, David and Dowling, Steve and Dunning, Sheila and Ecoffet, Adrien and Eleti, Atty and Eloundou, Tyna and Farhi, David and Fedus, Liam and Felix, Niko and Fishman, Sim{\'o}n Posada and Forte, Juston and Fulford, Isabella and Gao, Leo and Georges, Elie and Gibson, Christian and Goel, Vik and Gogineni, Tarun and Goh, Gabriel and Gontijo-Lopes, Rapha and Gordon, Jonathan and Grafstein, Morgan and Gray, Scott and Greene, Ryan and Gross, Joshua and Gu, Shixiang Shane and Guo, Yufei and Hallacy, Chris and Han, Jesse and Harris, Jeff and He, Yuchen and Heaton, Mike and Heidecke, Johannes and Hesse, Chris and Hickey, Alan and Hickey, Wade and Hoeschele, Peter and Houghton, Brandon and Hsu, Kenny and Hu, Shengli and Hu, Xin and Huizinga, Joost and Jain, Shantanu and Jain, Shawn and Jang, Joanne and Jiang, Angela and Jiang, Roger and Jin, Haozhun and Jin, Denny and Jomoto, Shino and Jonn, Billie and Jun, Heewoo and Kaftan, Tomer and Kaiser, {\L}ukasz and Kamali, Ali and Kanitscheider, Ingmar and Keskar, Nitish Shirish and Khan, Tabarak and Kilpatrick, Logan and Kim, Jong Wook and Kim, Christina and Kim, Yongjik and Kirchner, Jan Hendrik and Kiros, Jamie and Knight, Matt and Kokotajlo, Daniel and Kondraciuk, {\L}ukasz and Kondrich, Andrew and Konstantinidis, Aris and Kosic, Kyle and Krueger, Gretchen and Kuo, Vishal and Lampe, Michael and Lan, Ikai and Lee, Teddy and Leike, Jan and Leung, Jade and Levy, Daniel and Li, Chak Ming and Lim, Rachel and Lin, Molly and Lin, Stephanie and Litwin, Mateusz and Lopez, Theresa and Lowe, Ryan and Lue, Patricia and Makanju, Anna and Malfacini, Kim and Manning, Sam and Markov, Todor and Markovski, Yaniv and Martin, Bianca and Mayer, Katie and Mayne, Andrew and McGrew, Bob and McKinney, Scott Mayer and McLeavey, Christine and McMillan, Paul and McNeil, Jake and Medina, David and Mehta, Aalok and Menick, Jacob and Metz, Luke and Mishchenko, Andrey and Mishkin, Pamela and Monaco, Vinnie and Morikawa, Evan and Mossing, Daniel and Mu, Tong and Murati, Mira and Murk, Oleg and M{\'e}ly, David and Nair, Ashvin and Nakano, Reiichiro and Nayak, Rajeev and Neelakantan, Arvind and Ngo, Richard and Noh, Hyeonwoo and Ouyang, Long and O'Keefe, Cullen and Pachocki, Jakub and Paino, Alex and Palermo, Joe and Pantuliano, Ashley and Parascandolo, Giambattista and Parish, Joel and Parparita, Emy and Passos, Alex and Pavlov, Mikhail and Peng, Andrew and Perelman, Adam and Peres, Filipe de Avila Belbute and Petrov, Michael and Pinto, Henrique Ponde de Oliveira and Pokorny, Michael and Pokrass, Michelle and Pong, Vitchyr H. and Powell, Tolly and Power, Alethea and Power, Boris and Proehl, Elizabeth and Puri, Raul and Radford, Alec and Rae, Jack and Ramesh, Aditya and Raymond, Cameron and Real, Francis and Rimbach, Kendra and Ross, Carl and Rotsted, Bob and Roussez, Henri and Ryder, Nick and Saltarelli, Mario and Sanders, Ted and Santurkar, Shibani and Sastry, Girish and Schmidt, Heather and Schnurr, David and Schulman, John and Selsam, Daniel and Sheppard, Kyla and Sherbakov, Toki and Shieh, Jessica and Shoker, Sarah and Shyam, Pranav and Sidor, Szymon and Sigler, Eric and Simens, Maddie and Sitkin, Jordan and Slama, Katarina and Sohl, Ian and Sokolowsky, Benjamin and Song, Yang and Staudacher, Natalie and Such, Felipe Petroski and Summers, Natalie and Sutskever, Ilya and Tang, Jie and Tezak, Nikolas and Thompson, Madeleine B. and Tillet, Phil and Tootoonchian, Amin and Tseng, Elizabeth and Tuggle, Preston and Turley, Nick and Tworek, Jerry and Uribe, Juan Felipe Cer{\'o}n and Vallone, Andrea and Vijayvergiya, Arun and Voss, Chelsea and Wainwright, Carroll and Wang, Justin Jay and Wang, Alvin and Wang, Ben and Ward, Jonathan and Wei, Jason and Weinmann, CJ and Welihinda, Akila and Welinder, Peter and Weng, Jiayi and Weng, Lilian and Wiethoff, Matt and Willner, Dave and Winter, Clemens and Wolrich, Samuel and Wong, Hannah and Workman, Lauren and Wu, Sherwin and Wu, Jeff and Wu, Michael and Xiao, Kai and Xu, Tao and Yoo, Sarah and Yu, Kevin and Yuan, Qiming and Zaremba, Wojciech and Zellers, Rowan and Zhang, Chong and Zhang, Marvin and Zhao, Shengjia and Zheng, Tianhao and Zhuang, Juntang and Zhuk, William and Zoph, Barret},
  journal={arXiv preprint arXiv:2303.08774},
  year={2023}
}

@article{grattafiori2024llama,
  title={The llama 3 herd of models},
  author    = {Grattafiori, Aaron and Dubey, Abhimanyu and Jauhri, Abhinav and Pandey, Abhinav and Kadian, Abhishek and Al-Dahle, Ahmad and Letman, Aiesha and Mathur, Akhil and Schelten, Alan and Vaughan, Alex and Yang, Amy and Fan, Angela and Goyal, Anirudh and Hartshorn, Anthony and Yang, Aobo and Mitra, Archi and Sravankumar, Archie and Korenev, Artem and Hinsvark, Arthur and Rao, Arun and Zhang, Aston and Rodriguez, Aurelien and Gregerson, Austen and Spataru, Ava and Roziere, Baptiste and Biron, Bethany and Tang, Binh and Chern, Bobbie and Caucheteux, Charlotte and Nayak, Chaya and Bi, Chloe and Marra, Chris and McConnell, Chris and Keller, Christian and Touret, Christophe and Wu, Chunyang and Wong, Corinne and Ferrer, Cristian Canton and Nikolaidis, Cyrus and Allonsius, Damien and Song, Daniel and Pintz, Danielle and Livshits, Danny and Wyatt, Danny and Esiobu, David and Choudhary, Dhruv and Mahajan, Dhruv and Garcia-Olano, Diego and Perino, Diego and Hupkes, Dieuwke and Lakomkin, Egor and AlBadawy, Ehab and Lobanova, Elina and Dinan, Emily and Smith, Eric Michael and Radenovic, Filip and Guzm{\'a}n, Francisco and Zhang, Frank and Synnaeve, Gabriel and Lee, Gabrielle and Anderson, Georgia Lewis and Thattai, Govind and Nail, Graeme and Mialon, Gregoire and Pang, Guan and Cucurell, Guillem and Nguyen, Hailey and Korevaar, Hannah and Xu, Hu and Touvron, Hugo and Zarov, Iliyan and Ibarra, Imanol Arrieta and Kloumann, Isabel and Misra, Ishan and Evtimov, Ivan and Zhang, Jack and Copet, Jade and Lee, Jaewon and Geffert, Jan and Vranes, Jana and Park, Jason and Mahadeokar, Jay and Shah, Jeet and van der Linde, Jelmer and Billock, Jennifer and Hong, Jenny and Lee, Jenya and Fu, Jeremy and Chi, Jianfeng and Huang, Jianyu and Liu, Jiawen and Wang, Jie and Yu, Jiecao and Bitton, Joanna and Spisak, Joe and Park, Jongsoo and Rocca, Joseph and Johnstun, Joshua and Saxe, Joshua and Jia, Junteng and Alwala, Kalyan Vasuden and Prasad, Karthik and Upasani, Kartikeya and Plawiak, Kate and Li, Ke and Heafield, Kenneth and Stone, Kevin and El-Arini, Khalid and Iyer, Krithika and Malik, Kshitiz and Chiu, Kuenley and Bhalla, Kunal and Lakhotia, Kushal and Rantala-Yeary, Lauren and van der Maaten, Laurens and Chen, Lawrence and Tan, Liang and Jenkins, Liz and Martin, Louis and Madaan, Lovish and Malo, Lubo and Blecher, Lukas and Landzaat, Lukas and de Oliveira, Luke and Muzzi, Madeline and Pasupuleti, Mahesh and Singh, Mannat and Paluri, Manohar and Kardas, Marcin and Tsimpoukelli, Maria and Oldham, Mathew and Rita, Mathieu and Pavlova, Maya and Kambadur, Melanie and Lewis, Mike and Si, Min and Singh, Mitesh Kumar and Hassan, Mona and Goyal, Naman and Torabi, Narjes and Bashlykov, Nikolay and Bogoychev, Nikolay and Chatterji, Niladri and Zhang, Ning and Duchenne, Olivier and {\c C}elebi, Onur and Alrassy, Patrick and Zhang, Pengchuan and Li, Pengwei and Vasic, Petar and Weng, Peter and Bhargava, Prajjwal and Dubal, Pratik and Krishnan, Praveen and Koura, Punit Singh and Xu, Puxin and He, Qing and Dong, Qingxiao and Srinivasan, Ragavan and Ganapathy, Raj and Calderer, Ramon and Cabral, Ricardo Silveira and Stojnic, Robert and Raileanu, Roberta and Maheswari, Rohan and Girdhar, Rohit and Patel, Rohit and Sauvestre, Romain and Polidoro, Ronnie and Sumbaly, Roshan and Taylor, Ross and Silva, Ruan and Hou, Rui and Wang, Rui and Hosseini, Saghar and Chennabasappa, Sahana and Singh, Sanjay and Bell, Sean and Kim, Seohyun Sonia and Edunov, Sergey and Nie, Shaoliang and Narang, Sharan and Raparthy, Sharath and Shen, Sheng and Wan, Shengye and Bhosale, Shruti and Zhang, Shun and Vandenhende, Simon and Batra, Soumya and Whitman, Spencer and Sootla, Sten and Collot, Stephane and Gururangan, Suchin and Borodinsky, Sydney and Herman, Tamar and Fowler, Tara and Sheasha, Tarek and Georgiou, Thomas and Scialom, Thomas and Speckbacher, Tobias and Mihaylov, Todor and Xiao, Tong and Karn, Ujjwal and Goswami, Vedanuj and Gupta, Vibhor and Ramanathan, Vignesh and Kerkez, Viktor and Gonguet, Vincent and Do, Virginie and Vogeti, Vish and Albiero, V{\'i}tor and Petrovic, Vladan and Chu, Weiwei and Xiong, Wenhan and Fu, Wenyin and Meers, Whitney and Martinet, Xavier and Wang, Xiaodong and Wang, Xiaofang and Tan, Xiaoqing Ellen and Xia, Xide and Xie, Xinfeng and Jia, Xuchao and Wang, Xuewei and Goldschlag, Yaelle and Gaur, Yashesh and Babaei, Yasmine and Wen, Yi and Song, Yiwen and Zhang, Yuchen and Li, Yue and Mao, Yuning and Coudert, Zacharie Delpierre and Yan, Zheng and Chen, Zhengxing and Papakipos, Zoe and Singh, Aaditya and Srivastava, Aayushi and Jain, Abha and Kelsey, Adam and Shajnfeld, Adam and Gangidi, Adithya and Victoria, Adolfo and Goldstand, Ahuva and Menon, Ajay and Sharma, Ajay and Boesenberg, Alex and Baevski, Alexei and Feinstein, Allie and Kallet, Amanda and Sangani, Amit and Teo, Amos and Yunus, Anam and Lupu, Andrei and Alvarado, Andres and Caples, Andrew and Gu, Andrew and Ho, Andrew and Poulton, Andrew and Ryan, Andrew and Ramchandani, Ankit and Dong, Annie and Franco, Annie and Goyal, Anuj and Saraf, Aparajita and Chowdhury, Arkabandhu and Gabriel, Ashley and Bharambe, Ashwin and Eisenman, Assaf and Yazdan, Azadeh and James, Beau and Maurer, Ben and Leonhardi, Benjamin and Huang, Bernie and Loyd, Beth and De Paola, Beto and Paranjape, Bhargavi and Liu, Bing and Wu, Bo and Ni, Boyu and Hancock, Braden and Wasti, Bram and Spence, Brandon and Stojkovic, Brani and Gamido, Brian and Montalvo, Britt and Parker, Carl and Burton, Carly and Mejia, Catalina and Liu, Ce and Wang, Changhan and Kim, Changkyu and Zhou, Chao and Hu, Chester and Chu, Ching-Hsiang and Cai, Chris and Tindal, Chris and Feichtenhofer, Christoph and Gao, Cynthia and Civin, Damon and Beaty, Dana and Kreymer, Daniel and Li, Daniel and Adkins, David and Xu, David and Testuggine, Davide and David, Delia and Parikh, Devi and Liskovich, Diana and Foss, Didem and Wang, Dingkang and Le, Duc and Holland, Dustin and Dowling, Edward and Jamil, Eissa and Montgomery, Elaine and Presani, Eleonora and Hahn, Emily and Wood, Emily and Le, Eric-Tuan and Brinkman, Erik and Arcaute, Esteban and Dunbar, Evan and Smothers, Evan and Sun, Fei and Kreuk, Felix and Tian, Feng and Kokkinos, Filippos and Ozgenel, Firat and Caggioni, Francesco and Kanayet, Frank and Seide, Frank and Florez, Gabriela Medina and Schwarz, Gabriella and Badeer, Gada and Swee, Georgia and Halpern, Gil and Herman, Grant and Sizov, Grigory and Zhang, Guangyi and Lakshminarayanan, Guna and Inan, Hakan and Shojanazeri, Hamid and Zou, Han and Wang, Hannah and Zha, Hanwen and Habeeb, Haroun and Rudolph, Harrison and Suk, Helen and Aspegren, Henry and Goldman, Hunter and Zhan, Hongyuan and Damlaj, Ibrahim and Molybog, Igor and Tufanov, Igor and Leontiadis, Ilias and Veliche, Irina-Elena and Gat, Itai and Weissman, Jake and Geboski, James and Kohli, James and Lam, Janice and Asher, Japhet and Gaya, Jean-Baptiste and Marcus, Jeff and Tang, Jeff and Chan, Jennifer and Zhen, Jenny and Reizenstein, Jeremy and Teboul, Jeremy and Zhong, Jessica and Jin, Jian and Yang, Jingyi and Cummings, Joe and Carvill, Jon and Shepard, Jon and McPhie, Jonathan and Torres, Jonathan and Ginsburg, Josh and Wang, Junjie and Wu, Kai and U, Kam Hou and Saxena, Karan and Khandelwal, Kartikay and Zand, Katayoun and Matosich, Kathy and Veeraraghavan, Kaushik and Michelena, Kelly and Li, Keqian and Jagadeesh, Kiran and Huang, Kun and Chawla, Kunal and Huang, Kyle and Chen, Lailin and Garg, Lakshya and A, Lavender and Silva, Leandro and Bell, Lee and Zhang, Lei and Guo, Liangpeng and Yu, Licheng and Moshkovich, Liron and Wehrstedt, Luca and Khabsa, Madian and Avalani, Manav and Bhatt, Manish and Mankus, Martynas and Hasson, Matan and Lennie, Matthew and Reso, Matthias and Groshev, Maxim and Naumov, Maxim and Lathi, Maya and Keneally, Meghan and Liu, Miao and Seltzer, Michael L. and Valko, Michal and Restrepo, Michelle and Patel, Mihir and Vyatskov, Mik and Samvelyan, Mikayel and Clark, Mike and Macey, Mike and Wang, Mike and Hermoso, Miquel Jubert and Metanat, Mo and Rastegari, Mohammad and Bansal, Munish and Santhanam, Nandhini and Parks, Natascha and White, Natasha and Bawa, Navyata and Singhal, Nayan and Egebo, Nick and Usunier, Nicolas and Mehta, Nikhil and Laptev, Nikolay Pavlovich and Dong, Ning and Cheng, Norman and Chernoguz, Oleg and Hart, Olivia and Salpekar, Omkar and Kalinli, Ozlem and Kent, Parkin and Parekh, Parth and Saab, Paul and Balaji, Pavan and Rittner, Pedro and Bontrager, Philip and Roux, Pierre and Dollar, Piotr and Zvyagina, Polina and Ratanchandani, Prashant and Yuvraj, Pritish and Liang, Qian and Alao, Rachad and Rodriguez, Rachel and Ayub, Rafi and Murthy, Raghotham and Nayani, Raghu and Mitra, Rahul and Parthasarathy, Rangaprabhu and Li, Raymond and Hogan, Rebekkah and Battey, Robin and Wang, Rocky and Howes, Russ and Rinott, Ruty and Mehta, Sachin and Siby, Sachin and Bondu, Sai Jayesh and Datta, Samyak and Chugh, Sara and Hunt, Sara and Dhillon, Sargun and Sidorov, Sasha and Pan, Satadru and Mahajan, Saurabh and Verma, Saurabh and Yamamoto, Seiji and Ramaswamy, Sharadh and Lindsay, Shaun and Lindsay, Shaun and Feng, Sheng and Lin, Shenghao and Zha, Shengxin Cindy and Patil, Shishir and Shankar, Shiva and Zhang, Shuqiang and Zhang, Shuqiang and Wang, Sinong and Agarwal, Sneha and Sajuyigbe, Soji and Chintala, Soumith and Max, Stephanie and Chen, Stephen and Kehoe, Steve and Satterfield, Steve and Govindaprasad, Sudarshan and Gupta, Sumit and Deng, Summer and Cho, Sungmin and Virk, Sunny and Subramanian, Suraj and Choudhury, Sy and Goldman, Sydney and Remez, Tal and Glaser, Tamar and Best, Tamara and Koehler, Thilo and Robinson, Thomas and Li, Tianhe and Zhang, Tianjun and Matthews, Tim and Chou, Timothy and Shaked, Tzook and Vontimitta, Varun and Ajayi, Victoria and Montanez, Victoria and Mohan, Vijai and Kumar, Vinay Satish and Mangla, Vishal and Ionescu, Vlad and Poenaru, Vlad and Mihailescu, Vlad Tiberiu and Ivanov, Vladimir and Li, Wei and Wang, Wenchen and Jiang, Wenwen and Bouaziz, Wes and Constable, Will and Tang, Xiaocheng and Wu, Xiaojian and Wang, Xiaolan and Wu, Xilun and Gao, Xinbo and Kleinman, Yaniv and Chen, Yanjun and Hu, Ye and Jia, Ye and Qi, Ye and Li, Yenda and Zhang, Yilin and Zhang, Ying and Adi, Yossi and Nam, Youngjin and Wang, Yu and Zhao, Yu and Hao, Yuchen and Qian, Yundi and Li, Yunlu and He, Yuzi and Rait, Zach and DeVito, Zachary and Rosnbrick, Zef and Wen, Zhaoduo and Yang, Zhenyu and Zhao, Zhiwei and Ma, Zhiyu},
  journal={arXiv preprint arXiv:2407.21783},
  year={2024}
}

@article{pope2023efficiently,
  title={Efficiently scaling transformer inference},
  author={Pope, Reiner and Douglas, Sholto and Chowdhery, Aakanksha and Devlin, Jacob and Bradbury, James and Heek, Jonathan and Xiao, Kefan and Agrawal, Shivani and Dean, Jeff},
  journal={Proceedings of machine learning and systems},
  volume={5},
  pages={606--624},
  year={2023}
}

@article{agrawal2023sarathi,
  title={Sarathi: Efficient llm inference by piggybacking decodes with chunked prefills},
  author={Agrawal, Amey and Panwar, Ashish and Mohan, Jayashree and Kwatra, Nipun and Gulavani, Bhargav S and Ramjee, Ramachandran},
  journal={arXiv preprint arXiv:2308.16369},
  year={2023}
}

@article{qin2024mooncake,
  title={Mooncake: A kvcache-centric disaggregated architecture for llm serving},
  author    = {Qin, Ruoyu and Li, Zheming and He, Weiran and Cui, Jialei and Tang, Heyi and Ren, Feng and Ma, Teng and Cai, Shangming and Zhang, Yineng and Zhang, Mingxing and Wu, Yongwei and Zheng, Weimin and Xu, Xinran},
  journal={ACM Transactions on Storage},
  year={2024},
  publisher={ACM New York, NY}
}

@article{kung1982systolic,
  title={Why systolic architecture?},
  author    = {Kung, Hsiang-Tsung},
  year={1982},
  publisher={Design Research Center, Carnegie-Mellon University Pittsburgh, PA, USA}
}

@article{chen2016eyeriss,
  title={Eyeriss: An energy-efficient reconfigurable accelerator for deep convolutional neural networks},
  author={Chen, Yu-Hsin and Krishna, Tushar and Emer, Joel S and Sze, Vivienne},
  journal={IEEE journal of solid-state circuits},
  volume={52},
  number={1},
  pages={127--138},
  year={2016},
  publisher={IEEE}
}

@inproceedings{jouppi2017datacenter,
  title={In-datacenter performance analysis of a tensor processing unit},
  author    = {Jouppi, Norman P. and Young, Cliff and Patil, Nishant and Patterson, David and Agrawal, Gaurav and Bajwa, Raminder and Bates, Sarah and Bhatia, Suresh and Boden, Nan and Borchers, Al and Boyle, Rick and Cantin, Pierre-luc and Chao, Clifford and Clark, Chris and Coriell, Jeremy and Daley, Mike and Dau, Matt and Dean, Jeffrey and Gelb, Ben and Ghaemmaghami, Tara Vazir and Gottipati, Rajendra and Gulland, William and Hagmann, Robert and Ho, C. Richard and Hogberg, Doug and Hu, John and Hundt, Robert and Hurt, Dan and Ibarz, Julian and Jaffey, Aaron and Jaworski, Alek and Kaplan, Alexander and Khaitan, Harshit and Koch, Andy and Kumar, Naveen and Lacy, Steve and Laudon, James and Law, James and Le, Diemthu and Leary, Chris and Liu, Zhuyuan and Lucke, Kyle and Lundin, Alan and MacKean, Gordon and Maggiore, Adriana and Mahony, Maire and Miller, Kieran and Nagarajan, Rahul and Narayanaswami, Ravi and Ni, Ray and Nix, Kathy and Norrie, Thomas and Omernick, Mark and Penukonda, Narayana and Phelps, Andy and Ross, Jonathan and Ross, Matt and Salek, Amir and Samadiani, Emad and Severn, Chris and Sizikov, Gregory and Snelham, Matthew and Souter, Jed and Steinberg, Dan and Swing, Andy and Tan, Mercedes and Thorson, Gregory and Tian, Bo and Toma, Horia and Tuttle, Erick and Vasudevan, Vijay and Walter, Richard and Wang, Walter and Wilcox, Eric and Yoon, Doe Hyun},
  booktitle={Proceedings of the 44th annual international symposium on computer architecture},
  pages={1--12},
  year={2017}
}

@inproceedings{jouppi2023tpu,
  title={Tpu v4: An optically reconfigurable supercomputer for machine learning with hardware support for embeddings},
  author    = {Jouppi, Norman P. and Kurian, George and Li, Sheng and Ma, Peter and Nagarajan, Rahul and Nai, Lifeng and Patil, Nishant and Subramanian, Suvinay and Swing, Andy and Towles, Brian and Young, Cliff and Zhou, Xiang and Zhou, Zongwei and Patterson, David},
  booktitle={Proceedings of the 50th annual international symposium on computer architecture},
  pages={1--14},
  year={2023}
}

@article{shao2023omniquant,
  title={Omniquant: Omnidirectionally calibrated quantization for large language models},
  author={Shao, Wenqi and Chen, Mengzhao and Zhang, Zhaoyang and Xu, Peng and Zhao, Lirui and Li, Zhiqian and Zhang, Kaipeng and Gao, Peng and Qiao, Yu and Luo, Ping},
  journal={arXiv preprint arXiv:2308.13137},
  year={2023}
}

@article{kim2023squeezellm,
  title={Squeezellm: Dense-and-sparse quantization},
  author={Kim, Sehoon and Hooper, Coleman and Gholami, Amir and Dong, Zhen and Li, Xiuyu and Shen, Sheng and Mahoney, Michael W and Keutzer, Kurt},
  journal={arXiv preprint arXiv:2306.07629},
  year={2023}
}

@article{ashkboos2024quarot,
  title={Quarot: Outlier-free 4-bit inference in rotated llms},
  author={Ashkboos, Saleh and Mohtashami, Amirkeivan and Croci, Maximilian L and Li, Bo and Cameron, Pashmina and Jaggi, Martin and Alistarh, Dan and Hoefler, Torsten and Hensman, James},
  journal={Advances in Neural Information Processing Systems},
  volume={37},
  pages={100213--100240},
  year={2024}
}

@inproceedings{clark2019boolq,
  title={Boolq: Exploring the surprising difficulty of natural yes/no questions},
  author={Clark, Christopher and Lee, Kenton and Chang, Ming-Wei and Kwiatkowski, Tom and Collins, Michael and Toutanova, Kristina},
  booktitle={Proceedings of the 2019 conference of the north American chapter of the association for computational linguistics: Human language technologies, volume 1 (long and short papers)},
  pages={2924--2936},
  year={2019}
}

@inproceedings{zellers2019hellaswag,
  title={Hellaswag: Can a machine really finish your sentence?},
  author={Zellers, Rowan and Holtzman, Ari and Bisk, Yonatan and Farhadi, Ali and Choi, Yejin},
  booktitle={Proceedings of the 57th annual meeting of the association for computational linguistics},
  pages={4791--4800},
  year={2019}
}

@article{cobbe2021training,
  title={Training verifiers to solve math word problems},
  author    = {Cobbe, Karl and Kosaraju, Vineet and Bavarian, Mohammad and Chen, Mark and Jun, Heewoo and Kaiser, Lukasz and Plappert, Matthias and Tworek, Jerry and Hilton, Jacob and Nakano, Reiichiro and Hesse, Christopher and Schulman, John},
  journal={arXiv preprint arXiv:2110.14168},
  year={2021}
}

@inproceedings{bisk2020piqa,
  title={Piqa: Reasoning about physical commonsense in natural language},
  author    = {Bisk, Yonatan and Zellers, Rowan and Le Bras, Ronan and Gao, Jianfeng and Choi, Yejin},
  booktitle={Proceedings of the AAAI conference on artificial intelligence},
  volume={34},
  number={05},
  pages={7432--7439},
  year={2020}
}

@article{clark2018think,
  title={Think you have solved question answering? try arc, the ai2 reasoning challenge},
  author={Clark, Peter and Cowhey, Isaac and Etzioni, Oren and Khot, Tushar and Sabharwal, Ashish and Schoenick, Carissa and Tafjord, Oyvind},
  journal={arXiv preprint arXiv:1803.05457},
  year={2018}
}

@inproceedings{ghodrati2020planaria,
  title={Planaria: Dynamic architecture fission for spatial multi-tenant acceleration of deep neural networks},
  author    = {Ghodrati, Soroush and Ahn, Byung Hoon and Kim, Joon Kyung and Kinzer, Sean and Yatham, Brahmendra Reddy and Alla, Navateja and Sharma, Hardik and Alian, Mohammad and Ebrahimi, Eiman and Kim, Nam Sung and Young, Cliff and Esmaeilzadeh, Hadi},
  booktitle={2020 53rd Annual IEEE/ACM International Symposium on Microarchitecture (MICRO)},
  pages={681--697},
  year={2020},
  organization={IEEE}
}
